\documentclass[structabstract]{aa}
\usepackage{graphicx}
\usepackage{txfonts}
\usepackage{longtable}
\newcommand{\tablenotea}[1]{\parbox{9.8cm}{ \indent \footnotesize{\textsc{Notes.--}~#1}}}
\newcommand{\tablenoteb}[1]{\parbox{8.7cm}{ \indent \footnotesize{\textsc{}~#1}}}
\newcommand{\tablenotec}[1]{\parbox{8.6cm}{ \indent \footnotesize{\textsc{}~#1}}}
\newcommand{\tablenoted}[1]{\parbox{8.6cm}{ \indent \footnotesize{\textsc{}~#1}}}
\newcommand{\jms}{J. Mol. Spectr.}                     
\newcommand{\jmst}{J. Mol. Struct.}                    
\newcommand{\pccp}{Phys. Chem. Chem. Phys.}            
\newcommand{\jpbamop}{J. Phys. B: At. Mol. Opt. Phys.} 
\newcommand{\zfn}{Z. Naturforsch.}                     
\newcommand{\fdis}{Faraday Discuss.}                   
\newcommand{\nature}{Nature}                           

\begin{document}
\title{Molecular abundances in the inner layers of IRC +10216\thanks{Based on
observations carried out with the IRAM 30-meter telescope. IRAM is
supported by INSU/CNRS (France), MPG (Germany), and IGN (Spain).}}
\titlerunning{Molecular abundances in the inner layers of IRC
+10216}
\authorrunning{Ag\'undez et al.}
\author{Marcelino Ag\'undez\inst{1,2,3}, Jos\'e Pablo
Fonfr\'ia\inst{4}, Jos\'e Cernicharo\inst{1}, Claudine
Kahane\inst{5}, Fabien Daniel\inst{1}, and Michel
Gu\'elin\inst{6}} \institute{Departamento de Astrof\'isica, Centro
de Astrobiolog\'ia, CSIC--INTA, Ctra. de Torrej\'on a Ajalvir km
4, 28850 Madrid, Spain \and Univ. Bordeaux, LAB, UMR 5804,
F-33270, Floirac, France;
\email{Marcelino.Agundez@obs.u-bordeaux1.fr} \and CNRS, LAB, UMR
5804, F-33270, Floirac, France \and Departamento de Estrellas y
Medio Interestelar, Instituto de Astronom\'ia, Universidad
Nacional Aut\'onoma de M\'exico, Ciudad Universitaria, 04510
M\'exico City, M\'exico \and Institut de Plan\'etologie et
d'Astrophysique de Grenoble (IPAG), Univ. J. Fourier and CNRS (UMR
5274), BP 53 F--38041 Grenoble C\'edex 9, France \and Institut de
Radioastronomie Millim\'etrique, 300 rue de la Piscine, 38406
Saint Martin d'H\'eres, France}

\date{Received; accepted}


\abstract
{The inner layers of circumstellar envelopes around AGB stars are
sites where a variety of processes such as thermochemical
equilibrium, shocks induced by the stellar pulsation, and
condensation of dust grains determine the chemical composition of
the material that is expelled into the outer envelope layers and,
ultimately, into interstellar space.}
{We aim at studying the abundances, throughout the whole
circumstellar envelope of the carbon star IRC +10216, of several
molecules formed in the inner layers in order to constrain the
different processes at work in such regions.}
{Observations towards IRC +10216 of CS, SiO, SiS, NaCl, KCl, AlCl,
AlF, and NaCN have been carried out with the IRAM 30-m telescope
in the 80--357.5 GHz frequency range. A large number of rotational
transitions covering a wide range of energy levels, including
highly excited vibrational states, are detected in emission and
serve to trace different regions of the envelope. Radiative
transfer calculations based on the LVG formalism have been
performed to derive molecular abundances from the innermost out to
the outer layers. The excitation calculations include infrared
pumping to excited vibrational states and inelastic collisions,
for which up-to-date rate coefficients for rotational and, in some
cases, ro-vibrational transitions are used.}
{We find that in the inner layers CS, SiO, and SiS have abundances
relative to H$_2$ of 4 $\times$ 10$^{-6}$, 1.8 $\times$ 10$^{-7}$,
and 3 $\times$ 10$^{-6}$, respectively, and that CS and SiS have
significant lower abundances in the outer envelope, which implies
that they actively contribute to the formation of dust. Moreover,
in the inner layers, the amount of sulfur and silicon in gas phase
molecules is only 27 \% for S and 5.6 \% for Si, implying that
these elements have already condensed onto grains, most likely in
the form of MgS and SiC. Metal--bearing molecules lock up a
relatively small fraction of metals, although our results indicate
that NaCl, KCl, AlCl, AlF, and NaCN, despite their refractory
character, are not significantly depleted in the cold outer
layers. In these regions a few percent of the metals Na, K, and Al
survive in the gas phase, either in atomic or molecular form, and
are therefore available to participate in the gas phase chemistry
in the outer envelope.}
{}

\keywords{astrochemistry --- line: identification --- molecular
processes --- stars: AGB and post-AGB --- circumstellar matter ---
stars: individual (IRC +10216)}

\maketitle
%

\section{Introduction}

The carbon star envelope IRC +10216 is among the best studied
astronomical objects (\cite{mor1975} 1975; \cite{gla1996} 1996;
\cite{mau1999} 1999; \cite{cer2000} 2000; \cite{mil2000} 2000).
The central low--mass AGB star CW Leo, losing mass at a high rate
(1--4 $\times$ 10$^{-5}$ M$_{\odot}$ yr$^{-1}$), is embedded in a
dense and nearly spherical circumstellar envelope composed of
molecular gas and dust particles. Being a relatively nearby source
(110--170 pc), it is one of the brightest infrared objects and
richest molecular sources -- with more than 80 molecules detected
-- in the sky. IRC +10216 is the prototype of carbon star and a
challenging source for chemical models.

Molecules observed in IRC +10216 form either in the atmosphere of
the central AGB star, from where they are expelled (\emph{parent}
species) or \emph{in situ} in the outer envelope (\emph{daughter}
species). Interferometric maps show that molecular emission peaks
either on the star or farther out in a hollow shell of radius
10--20$''$ (\cite{luc1995} 1995; \cite{gue1997} 1997). The parent
species are formed in the dense and hot atmosphere at
thermochemical equilibrium (TE), although the formation of dust
grains, shocks, and the penetration of interstellar ultraviolet
photons through the clumpy envelope (\cite{agu2006} 2006;
\cite{che2006} 2006; \cite{agu2010} 2010) can make molecular
abundances to deviate from TE. The chemical composition in the
inner layers is key as this material, once incorporated in the
outflowing wind, constitutes the basis for the rich photochemistry
and ion-neutral chemistry that takes place in the outer envelope
layers.

Probing molecular abundances in the warm (1000--2000 K) and dense
(10$^8$--10$^{14}$ cm$^{-3}$) inner layers requires the
observation of high energy rotational or ro-vibrational lines that
lie in the submillimeter and infrared regions of the spectrum,
which can hardly be observed from the ground. Efforts to this end
were undertaken by \cite{kea1993} (1993) and \cite{boy1994}
(1994), who observed with ground-based telescopes absorption lines
due to ro-vibrational transitions of C$_2$H$_2$, CH$_4$, SiH$_4$,
CS, SiO, NH$_3$, and SiS near 10 and 13.5 $\mu$m. \cite{fon2008}
(2008) made a sensitive spectral survey in the 11--14 $\mu$m range
that unveiled a forest of ro-vibrational lines arising from
C$_2$H$_2$ and HCN. A recent interferometric spectral survey in
the 0.9 mm atmospheric window by \cite{pat2011} (2011) allowed to
detect a good number of narrow unidentified lines arising from the
innermost envelope, inside the acceleration region, that they
tentatively assign to molecular rotational transitions within
excited vibrational states.

Space missions now freely explore spectral regions not accessible
from the ground. ISO observed a full infrared scan from 2.4 to 197
$\mu$m, albeit with a low spatial and spectral resolution, that
probed the inner layers of IRC +10216 through the lines of
abundant molecules such as CO, C$_2$H$_2$, and HCN (\cite{cer1996}
1996, 1999). More recently, spectral surveys carried out with
\emph{Herschel} with a high angular resolution have detected many
high energy rotational transitions of molecules such as CO, HCN,
CS, SiS, SiO, and SiC$_2$, at low spectral resolution using SPIRE
and PACS in the 55-672 $\mu$m range (\cite{dec2010a} 2010a), and
at high spectral resolution in the 488--1901 GHz range covered by
HIFI (\cite{cer2010} 2010).

Here we present an exhaustive analysis of the abundance and
excitation conditions of several molecules formed in the inner
layers of IRC +10216, based on high sensitive observations within
the full frequency coverage of the IRAM 30-m telescope (80-360
GHz). We focus on the parent molecules CS, SiO, SiS, NaCl, KCl,
AlCl, AlF, and NaCN. The cases of CO, HCN, and SiC$_2$, also
formed in the inner layers and reachable by the IRAM 30-m
telescope, have been treated recently (\cite{deb2012} 2012;
\cite{fon2008} 2008; \cite{cer2010} 2010, 2011). The observation
of a large number of rotational transitions covering a wide range
of excitation energies coupled to radiative transfer calculations
allows us to derive accurate abundances throughout the envelope.
Rotational transitions of CS, SiO, and SiS in highly excited
vibrational states are observed; for SiS up to $v$=5, with energy
levels above 3600 cm$^{-1}$ (5200 K). They put severe constraints
on the abundance of these species in the innermost regions.
Together with low energy transitions, the abundance can be tracked
from the inner to the outer layers, allowing to evaluate the
effects of shocks, grain formation, and gas phase chemistry.

\section{Astronomical observations}

The mm-wave observations of IRC +10216 on which the present study
is based were made with the IRAM 30-m telescope on Pico Veleta
(Spain). They cover the main atmospheric windows, at $\lambda$3 mm
(80--116 GHz), $\lambda$2 mm (129--183.5), and $\lambda$1 mm
(197--357.5). The observations at 3 and 2 mm were carried out
between 1986 and 2008 using the old {\footnotesize ABCD} SIS
receivers. A large part of the $\lambda$2 mm data have been
already presented as a spectral survey (\cite{cer2000} 2000). At
$\lambda$3 mm, the most sensitive observations were obtained after
2002 in the context of a coherent line survey not yet published
(Cernicharo et al. in preparation). Part of the spectra at 1 mm
were obtained from 1999 to 2008 using the {\footnotesize ABCD}
receivers although most of the data, in particular all spectra
with frequencies above 258 GHz, were observed from December 2009
to December 2010 with the new low--noise wide--band {\footnotesize
EMIR} receivers, in the context of a line survey (Kahane et al. in
preparation).

Both the old {\footnotesize ABCD} and the new {\footnotesize EMIR}
receivers are dual polarization and operate in single side band.
Image side band rejections are $>$20 dB at 3 mm and 2 mm, $\sim$10
dB at 1 mm with {\footnotesize ABCD} receivers, and $\geq$10 dB at
1 mm with {\footnotesize EMIR} receivers. Identification of image
side band lines during the observations with {\footnotesize ABCD}
receivers was done through shifts in the local oscillator
frequency. This was not necessary with {\footnotesize EMIR}
receivers because of the markedly different rejection of the
horizontal and vertical polarizations. The backends provided
spectral resolutions between 1 and 2 MHz. The coarser resolution
was used above 258 GHz where it corresponds to $\leq$2.3 km
s$^{-1}$, still good enough to resolve the lines in IRC +10216's
spectra which have typical widths of 20--30 km s$^{-1}$.

The observations were made in the wobbler switching observing mode
by nutating the secondary mirror by $\pm$90$''$ or $\pm$120$''$ at
a rate of 0.5 Hz. The intensity in Figures and Tables is expressed
in terms of $T_A^*$, the antenna temperature corrected for
atmospheric absorption and for antenna ohmic and spillover losses.
$T_A^*$ can be converted into $T_{\rm MB}$ (main beam brightness
temperature) by dividing by B$_{\rm eff}$/F$_{\rm eff}$, where
F$_{\rm eff}$, the telescope forward efficiency, is 0.95 at 3 mm,
0.93 at 2 mm, 0.91 at 1.3 mm, and 0.88 at 1 mm for {\footnotesize
ABCD} receivers, and 0.91 at 1.3 mm and 0.84 at 0.9 mm for
{\footnotesize EMIR} receivers, and B$_{\rm eff}$, the beam
efficiency, is given by $B_{\rm eff}$ = 0.828 $\times$ $\exp$
$\{-(\nu/341)^2\}$ for receivers {\footnotesize ABCD} and by
$B_{\rm eff}$ = 0.865 $\times$ $\exp$ $\{-(\nu/365)^2\}$ for
{\footnotesize EMIR} receivers, where $\nu$ is the frequency
expressed in GHz.

The pointing and focus of the telescope were checked every 1--2 h
on nearby planets or the quasar OJ 287. The error in the telescope
pointing is estimated to be 1--2$''$. The main beam of the IRAM
30-m telescope is given by HPBW($''$) = 2460/$\nu$(GHz), so that
it ranges from 30$''$ at 80 GHz to 7$''$ at 360 GHz.

IRC +10216 is a variable source with a period of 649 days and a
variation in the infrared (IR) flux from maximum to minimum phase
of 2.2 mag at 1.2 $\mu$m and 0.6 mag at 18 $\mu$m (\cite{leb1992}
1992). Since our observations were carried out over a time span of
several years, we checked for possible time variation in the
intensity of some lines, especially those that are radiatively
excited. Most of the studied lines in the $\lambda$3 and
$\lambda$2 mm bands were observed at various epochs corresponding
to different IR phases. As previously discussed in \cite{cer2000}
(2000), we have found no evidence of time variability of the line
intensities that exceeds the calibration uncertainties and could
be ascribed to changes of the IR flux. Calibration uncertainties,
mostly linked to corrections of atmospheric absorption and
pointing errors, are estimated to be $\leq$10 \% at 3 mm, $\leq$20
\% at 2 mm, and $\leq$30 \% at 1.3 mm. Most of the lines at
wavelengths shorter than 1.1 mm were observed just once, so that
no constraints can be put on their time variability.

Observations of CS cover the $J=2-1$ to $J=7-6$ rotational
transitions of the $v=0$ state, some rotational transitions of the
excited vibrational states $v=1,2,3$, as well as several
transitions of the rare isotopomers $^{13}$CS, C$^{34}$S, and
C$^{33}$S (see Table~\ref{table-cs-line-param}). For SiO, we
observed the $J=2-1$ through $J=8-7$ rotational transitions of the
$v=0$ state and 3 transitions of the $v=1$ state, as well as
several rotational transitions of the rare isotopomers $^{29}$SiO,
$^{30}$SiO, Si$^{18}$O, and Si$^{17}$O (see
Table~\ref{table-sio-line-param}). In the case of SiS, rotational
transitions from $J=5-4$ to $J=19-18$ are observed in vibrational
states up to $v=5$. A good number of lines of the ground and
excited vibrational states were also observed for the rare
isotopomers $^{29}$SiS, $^{30}$SiS, Si$^{34}$S, and Si$^{33}$S
(see Table~\ref{table-sis-line-param}). Finally, rotational
transitions of several metal--bearing molecules were observed in
their ground vibrational states, including the halides NaCl, KCl,
AlCl, and AlF, and the cyanide NaCN. For this latter, rotational
transitions with upper quantum numbers $J=5$ to $J=23$ and $K_a=0$
to $K_a=10$ were clearly detected, while in the cases of the
chlorine-containing molecules both the $^{35}$Cl and $^{37}$Cl
isotopomers were detected (see Tables~\ref{table-nacl-line-param},
\ref{table-kcl-line-param}, \ref{table-alcl-line-param},
\ref{table-alf-line-param}, and \ref{table-nacn-line-param}).

\onltab{1}{
\begin{table*} \caption{CS line parameters in IRC
+10216} \label{table-cs-line-param} \centering
\begin{tabular}{crrcll}
\hline \hline
\multicolumn{1}{c}{Transition} & \multicolumn{1}{c}{$\nu_{\rm rest}$} & \multicolumn{1}{c}{$\nu_{\rm obs}$} & \multicolumn{1}{c}{$\int$T$_{A}^*d$v} & \multicolumn{1}{c}{v$_{\rm exp}$} & \multicolumn{1}{c}{Notes} \\
                               & \multicolumn{1}{c}{(MHz)}            & \multicolumn{1}{c}{(MHz)}           & \multicolumn{1}{c}{K km s$^{-1}$}     & \multicolumn{1}{c}{km s$^{-1}$}   & \\
\hline
\multicolumn{6}{c}{CS} \\
$v$=0 J=2-1 &  97980.952 &  97980.8(1) & 117(3)  & 13.8(2)  & \\
$v$=0 J=3-2 & 146969.025 & 146968.6(2) & 267(6)  & 13.8(3)  & \\
$v$=0 J=4-3 & 195954.213 & 195953.7(4) & 286(8)  & 13.3(3)  & \\
$v$=0 J=5-4 & 244935.554 & 244935.1(3) & 308(8)  & 12.5(4)  & \\
$v$=0 J=6-5 & 293912.089 & 293911.7(3) & 423(10) & 12.6(4)  & \\
$v$=0 J=7-6 & 342882.854 & 342881.8(3) & 259(8)  & 12.3(4)  & \\
$v$=1 J=2-1 &  97270.997 &  97271.5(5) & 0.11(2) & 6.1(6)   & \$ $!$ \\
$v$=1 J=3-2 & 145904.089 & 145903.8(3) & 1.05(5) & 5.4(4)   & $\spadesuit$ \\
$v$=1 J=5-4 & 243160.648 & 243161.1(5) & 1.38(7) & 6.6(3)   & \$ \\
$v$=1 J=6-5 & 291782.190 & 291782.2(5) & 3.0(1)  & 11.8(8)  & \\
$v$=1 J=7-6 & 340397.957 & 340398.3(5) & 1.74(8) & 10.4(6)  & \\
$v$=2 J=6-5 & 289651.545 & 289651.9(5) & 0.60(5) & 5.8(6)   & \\
$v$=2 J=7-6 & 337912.189 & 337912.9(6) & 0.34(3) & 4.8(6)   & \\
$v$=3 J=5-4 & 239608.923 & 239610.1(5) & 0.07(1) & 3.3(2)   & \\
$v$=3 J=6-5 & 287520.096 & 287521.1(8) & 0.28(2) & 7.1(5)   & \\
$v$=3 J=7-6 & 335425.481 & 335425.8(8) & 0.14(2) & 3.4(5)   & \\
\hline
\multicolumn{6}{c}{$^{13}$CS} \\
$v$=0 J=2-1 &  92494.273 & 92494.2(2)  & 4.1(2)  & 14.2(8)  & \$ \\
$v$=0 J=3-2 & 138739.267 & 138739.1(2) & 10.7(5) & 14.1(4)  & \$ \\
$v$=0 J=5-4 & 231220.689 & 231220.6(2) & 16.8(8) & 14.2(3)  & \\
$v$=0 J=6-5 & 277455.403 & 277456.2(6) & 22.6(10)& 14.5(6)  & \\
$v$=0 J=7-6 & 323684.976 & 323686.3(10)& 12.5(8) & 14.4(8)  & \\
$v$=1 J=5-4 & 229592.986 & 229593.0(6) & 0.05(1) & 2.5(2)   & \\
\hline
\multicolumn{6}{c}{C$^{34}$S} \\
$v$=0 J=2-1 &  96412.956 &  96412.9(2) & 10.1(4) & 14.3(3)  & \\
$v$=0 J=3-2 & 144617.107 & 144617.0(1) & 21.5(5) & 13.8(2)  & \\
$v$=0 J=5-4 & 241016.098 & 241016.1(1) & 39.4(8) & 14.4(2)  & \\
$v$=0 J=6-5 & 289209.077 & 289208.9(5) & 40(1)   & 13.7(4)  & \\
$v$=0 J=7-6 & 337396.471 & 337396.1(5) & 24.2(8) & 14.0(4)  & \\
$v$=1 J=3-2 & 143577.671 & 143576.9(10)& 0.04(1) & 6.9(10)  & $!$ \\
$v$=1 J=6-5 & 287130.180 & 287131.1(5) & 0.18(2) & 4.9(7)   & \\
\hline
\multicolumn{6}{c}{C$^{33}$S} \\
$v$=0 J=2-1 &  97172.063 &  97171.6(5) & 1.71(8) & 15.1(7)  & \$ \\
$v$=0 J=3-2 & 145755.730 & 145755.6(4) & 4.9(2)  & 14.4(7)  & \$ \\
$v$=0 J=5-4 & 242913.608 & 242913.7(2) & 8.3(1)  & 14.4(4)  & \\
$v$=0 J=6-5 & 291485.927 & 291485.9(6) & 10.0(3) & 14.2(6)  & \$ \\
$v$=0 J=7-6 & 340052.572 & 340052.8(8) & 6.1(2)  & 13.6(8)  & \$ \\
\hline
\end{tabular}
\tablenotea{
Numbers in parentheses are 1$\sigma$ errors in units of the last digits. \\
Observed frequencies $\nu_{\rm obs}$ were derived adopting a systemic velocity $v_{\rm LSR}$ for IRC +10216 of -26.5 km s$^{-1}$ (\cite{cer2000} 2000). \\
$\|$         Line width parameter v$_{\rm exp}$ has been fixed. \\
\$           Blended with another line. An individual fit for each line is possible. \\
$\pounds$    Two transitions are very close in frequency. A single line is fitted. \\
$!$          Marginal detection. \\
$\spadesuit$ Clear or probable maser line. \\
$\infty$     Complex line profile. Line fit is not particularly good. \\
}
\end{table*}
}

\onltab{2}{
\begin{table*} \caption{SiO line parameters in IRC
+10216} \label{table-sio-line-param} \centering
\begin{tabular}{crrcll}
\hline \hline
\multicolumn{1}{c}{Transition} & \multicolumn{1}{c}{$\nu_{\rm rest}$} & \multicolumn{1}{c}{$\nu_{\rm obs}$} & \multicolumn{1}{c}{$\int$T$_{A}^*d$v} & \multicolumn{1}{c}{v$_{\rm exp}$} & \multicolumn{1}{c}{Notes} \\
                               & \multicolumn{1}{c}{(MHz)}            & \multicolumn{1}{c}{(MHz)}           & \multicolumn{1}{c}{K km s$^{-1}$}     & \multicolumn{1}{c}{km s$^{-1}$}   & \\
\hline
\multicolumn{6}{c}{SiO} \\
$v$=0 J=2-1 &  86846.971 & 86846.8(2)  & 52(2)   & 13.9(2)  & \\
$v$=0 J=3-2 & 130268.665 & 130268.5(2) & 106(4)  & 13.6(2)  & \\
$v$=0 J=4-3 & 173688.210 & 173687.6(3) & 153(6)  & 13.6(3)  & \\
$v$=0 J=5-4 & 217104.889 & 217104.6(2) & 145(7)  & 13.1(5)  & \\
$v$=0 J=6-5 & 260517.985 & 260517.8(2) & 161(8)  & 13.7(4)  & \\
$v$=0 J=7-6 & 303926.783 & 303926.6(2) & 173(8)  & 13.0(4)  & \\
$v$=0 J=8-7 & 347330.565 & 347329.9(7) & 153(10) & 13.0(7)  & \\
$v$=1 J=2-1 &  86243.430 & 86243.1(10) & 0.04(2) & 10.8(10) & \$ $!$ \\
$v$=1 J=5-4 & 215596.029 & 215596.0(5) & 0.54(5) & 13.4(8)  & \\
$v$=1 J=7-6 & 301814.369 & 301815.9(10)& 0.93(6) & 12.6(8)  & \\
\hline
\multicolumn{6}{c}{$^{29}$SiO} \\
$v$=0 J=2-1 &  85759.202 &  85759.0(2) & 4.2(2)  & 15.3(4)  & \\
$v$=0 J=4-3 & 171512.814 & 171513.0(2) & 11.9(5) & 14.1(3)  & \\
$v$=0 J=5-4 & 214385.778 & 214385.9(3) & 13.2(5) & 13.9(4)  & \\
$v$=0 J=6-5 & 257255.249 & 257255.7(5) & 11.1(5) & 14.0(5)  & \\
$v$=0 J=7-6 & 300120.528 & 300121.7(3) & 16.4(6) & 13.8(4)  & \\
$v$=0 J=8-7 & 342980.917 & 342980.9(5) & 11.2(5) & 13.4(5)  & \\
\hline
\multicolumn{6}{c}{$^{30}$SiO} \\
$v$=0 J=2-1 &  84746.193 &  84746.0(2) & 2.9(1)  & 14.6(3)  & \\
$v$=0 J=4-3 & 169486.930 & 169486.8(2) & 7.3(3)  & 14.5(3)  & \\
$v$=0 J=5-4 & 211853.546 & 211853.4(3) & 15.2(6) & 13.5(5)  & \\
$v$=0 J=6-5 & 254216.751 & 254216.7(6) & 9.5(5)  & 14.6(6)  & \\
$v$=0 J=7-6 & 296575.863 & 296575.9(4) & 10.4(2) & 14.0(5)  & \\
$v$=0 J=8-7 & 338930.199 & 338930.0(4) & 6.9(1)  & 14.0(5)  & \\
\hline
\multicolumn{6}{c}{Si$^{18}$O} \\
$v$=0 J=2-1 &  80704.907 & 80705.6(15) & 0.06(3) & 14.5(20) & $!$ \\
$v$=0 J=5-4 & 201751.443 & 201750.8(10)& 0.30(8) & 12.6(15) & $!$ \\
\hline
\multicolumn{6}{c}{Si$^{17}$O} \\
$v$=0 J=4-3 & 167171.974 & 167172.1(6)  & 0.49(3) & 14.3(6) & \\
$v$=0 J=5-4 & 208959.990 & 208960.4(8)  & 0.30(5) & 14.2(6) & \\
$v$=0 J=7-6 & 292525.404 & 292526.0(10) & 0.54(3) & 14.1(8) & \\
$v$=0 J=8-7 & 334301.474 & 334300.7(10) & 0.33(3) & 16.4(20)& \\
\hline
\end{tabular}
\end{table*}
}

\onllongtab{3}{
\begin{longtable}{lrrccl}
\caption{SiS line parameters in IRC +10216} \label{table-sis-line-param}\\
\hline \hline
\multicolumn{1}{c}{Transition} & \multicolumn{1}{c}{$\nu_{\rm rest}$} & \multicolumn{1}{c}{$\nu_{\rm obs}$} & \multicolumn{1}{c}{$\int$T$_{A}^*d$v} & \multicolumn{1}{c}{v$_{\rm exp}$} & \multicolumn{1}{c}{Notes} \\
                               & \multicolumn{1}{c}{(MHz)}            & \multicolumn{1}{c}{(MHz)}           & \multicolumn{1}{c}{K km s$^{-1}$}     & \multicolumn{1}{c}{km s$^{-1}$}   & \\
\hline
\endfirsthead
\caption{Continued.} \\
\hline
\multicolumn{1}{c}{Transition} & \multicolumn{1}{c}{$\nu_{\rm rest}$} & \multicolumn{1}{c}{$\nu_{\rm obs}$} & \multicolumn{1}{c}{$\int$T$_{A}^*d$v} & \multicolumn{1}{c}{v$_{\rm exp}$} & \multicolumn{1}{c}{Notes} \\
                               & \multicolumn{1}{c}{(MHz)}            & \multicolumn{1}{c}{(MHz)}           & \multicolumn{1}{c}{K km s$^{-1}$}     & \multicolumn{1}{c}{km s$^{-1}$}   & \\
\hline
\endhead
\hline
\endfoot
\hline
\endlastfoot
\multicolumn{6}{c}{SiS} \\
$v$=0 J=5-4   & 90771.564  & 90771.5(1)  & 98(3)   & 14.0(3)  & \\
$v$=0 J=6-5   & 108924.301 & 108924.3(1) & 89(3)   & 14.1(4)  & \\
$v$=0 J=8-7   & 145227.052 & 145227.0(1) & 211(6)  & 13.5(6)  & \\
$v$=0 J=9-8   & 163376.780 & 163376.5(1) & 255(7)  & 13.5(7)  & \\
$v$=0 J=10-9  & 181525.218 & 181525.0(2) & 178(5)  & 12.9(7)  & \\
$v$=0 J=11-10 & 199672.224 & 199672.3(10)& 238(10) & 16.1(15) & $\infty$ $\spadesuit$ \\
$v$=0 J=12-11 & 217817.653 & 217817.5(2) & 178(6)  & 13.4(8)  & \\
$v$=0 J=13-12 & 235961.363 & 235960.9(3) & 215(6)  & 12.9(8)  & \\
$v$=0 J=14-13 & 254103.211 & 254103.0(10)& 301(12) & 13.1(15) & $\infty$ $\spadesuit$ \\
$v$=0 J=15-14 & 272243.052 & 272242.6(10)& 371(14) & 12.5(15) & $\infty$ $\spadesuit$ \\
$v$=0 J=16-15 & 290380.744 & 290380.7(4) & 326(16) & 13.5(6)  & \\
$v$=0 J=17-16 & 308516.144 & 308515.3(6) & 185(8)  & 12.9(7)  & \\
$v$=0 J=18-17 & 326649.108 & 326648.1(8) & 232(13) & 13.0(8)  & \\
$v$=0 J=19-18 & 344779.492 & 344778.9(6) & 194(11) & 13.2(6)  & \\
$v$=1 J=5-4   & 90329.890  & 90329.9(2)  & 0.29(2) & 11.7(6)  & \\
$v$=1 J=6-5   & 108394.291 & 108394.7(1) & 0.68(4) & 9.7(6)   & \\
$v$=1 J=8-7   & 144520.367 & 144520.5(4) & 0.63(5) & 6.5(6)   & \\
$v$=1 J=9-8   & 162581.755 & 162580.4(10)& 1.24(10)& 9.8(10)  & \\
$v$=1 J=11-10 & 198700.517 & 198700.2(4) & 1.62(5) & 12.0(5)  & \\
$v$=1 J=12-11 & 216757.603 & 216757.7(4) & 2.08(6) & 12.9(6)  & \\
$v$=1 J=13-12 & 234812.968 & 234813.0(3) & 2.66(8) & 10.7(5)  & \\
$v$=1 J=14-13 & 252866.468 & 252868.1(10)& 2.93(10)& 12.3(10) & \\
$v$=1 J=15-14 & 270917.961 & 270916.9(7) & 3.5(1)  & 10.4(8)  & \\
$v$=1 J=16-15 & 288967.302 & 288967.5(6) & 6.5(2)  & 12.2(6)  & \\
$v$=1 J=19-18 & 343100.984 & 343100.4(6) & 3.7(1)  & 11.3(7)  & \\
$v$=2 J=5-4   & 89888.175  & 89888.4(3)  & 0.09(1) & 9.4(8)   & \\
$v$=2 J=6-5   & 107864.231 & 107864.1(2) & 0.19(2) & 10.4(7)  & \\
$v$=2 J=8-7   & 143813.615 & 143813.0(5) & 0.14(2) & 5.9(5)   & \\
$v$=2 J=11-10 & 197728.718 & 197728.7(6) & 0.37(4) & 10.4(9)  & \\
$v$=2 J=12-11 & 215697.452 & 215697.4(5) & 0.54(4) & 6.7(6)   & \\
$v$=2 J=13-12 & 233664.464 & 233663.6(10)& 0.40(6) & 6.1(10)  & \$ \\
$v$=2 J=14-13 & 251629.609 & 251628.6(8) & 1.42(8) & 9.6(8)   & \\
$v$=2 J=15-14 & 269592.744 & 269593.4(8) & 1.59(7) & 7.6(8)   & \\
$v$=2 J=16-15 & 287553.726 & 287551.6(10)& 0.44(4) & 8.6(10)  & \\
$v$=2 J=17-16 & 305512.410 & 305512.7(7) & 1.14(3) & 6.8(7)   & \\
$v$=2 J=19-18 & 341422.314 & 341422.4(7) & 0.51(3) & 8.0(8)   & \\
$v$=3 J=5-4   & 89446.404  & 89446.5(4)  & 0.023(7)& 3.5(10)  & \\
$v$=3 J=6-5   & 107334.104 & 107333.0(10)& 0.20(3) & 8.0(10)  & \\
$v$=3 J=8-7   & 143106.774 & 143107.3(8) & 0.11(2) & 4.1(8)   & \\
$v$=3 J=9-8   & 160991.456 & 160991.3(8) & 0.15(2) & 4.1(8)   & \\
$v$=3 J=10-9  & 178874.845 & 178875.4(8) & 0.12(2) & 3.7(7)   & \\
$v$=3 J=11-10 & 196756.796 & 196757.0(10)& 0.08(2) & 2.5(6)   & $!$ \\
$v$=3 J=12-11 & 214637.167 & 214636.4(8) & 0.18(2) & 7.2(6)   & \\
$v$=3 J=14-13 & 250392.591 & 250394.0(20)& 0.37(6) & 8.0(8)   & \\
$v$=3 J=15-14 & 268267.357 & 268267.5(10)& 0.28(4) & 4.6(10)  & \$ \\
$v$=3 J=16-15 & 286139.967 & 286140.4(8) & 0.38(4) & 5.0(8)   & \\
$v$=3 J=17-16 & 304010.278 & 304010.7(12)& 0.73(8) & 5.2(12)  & \$ \\
$v$=3 J=18-17 & 321878.146 & 321878.7(15)& 0.46(5) & 4.9(15)  & \$ \\
$v$=4 J=8-7   & 142399.819 & 142400.2(10)& 0.05(1) & 3.2(8)   & \\
$v$=4 J=12-11 & 213576.710 & 213576.9(5) & 0.21(4) & 3.5(5)   & \\
$v$=4 J=13-12 & 231366.976 & 231367.2(3) & 0.26(3) & 2.7(3)   & \\
$v$=4 J=14-13 & 249155.372 & 249155.6(4) & 0.17(4) & 3.5(4)   & \\
$v$=4 J=15-14 & 266941.754 & 266942.4(7) & 0.24(4) & 4.0(7)   & \$ \\
$v$=4 J=16-15 & 284725.978 & 284725.1(8) & 0.16(1) & 3.2(9)   & \\
$v$=4 J=17-16 & 302507.901 & 302509.3(15)& 0.28(4) & 5.0(15)  & \$ \\
$v$=4 J=18-17 & 320287.377 & 320286.4(20)& 0.09(2) & 3.3(10)  & $!$ \\
$v$=4 J=19-18 & 338064.264 & 338066.3(20)& 0.15(3) & 5.0(15)  & \\
$v$=4 J=20-19 & 355838.417 & 355839.9(15)& 0.22(3) & 7.3(15)  & \\
$v$=5 J=14-13 & 247917.909 & 247917.2(15)& 0.16(2) & 6.4(20)  & \\
$v$=5 J=16-15 & 283311.709 & 283313.1(20)& 0.16(2) & 2.9(10)  & \\
$v$=5 J=17-16 & 301005.224 & 301005.5(20)& 0.17(4) & 5.5(20)  & \$ \\
$v$=5 J=18-17 & 318696.291 & 318697.2(15)& 0.13(3) & 4.7(20)  & \$ \\
$v$=5 J=19-18 & 336384.765 & 336385.1(8) & 0.12(2) & 4.5(10)  & \\
\hline
              &            &             &         &          & \\
              &            &             &         &          & \\
\multicolumn{6}{c}{$^{29}$SiS} \\
$v$=0 J=5-4   & 89103.780  & 89103.8(1)  & 3.9(1)  & 14.8(5)  & \\
$v$=0 J=6-5   & 106923.018 & 106923.0(1) & 7.8(1)  & 15.1(5)  & \\
$v$=0 J=8-7   & 142558.872 & 142558.7(2) & 14.3(3) & 14.6(3)  & \\
$v$=0 J=9-8   & 160375.212 & 160375.1(1) & 19.0(4) & 14.0(3)  & \\
$v$=0 J=11-10 & 196004.027 & 196004.1(2) & 16.6(4) & 14.1(3)  & \\
$v$=0 J=12-11 & 213816.227 & 213816.1(2) & 20.1(4) & 14.0(3)  & \\
$v$=0 J=13-12 & 231626.770 & 231626.7(2) & 24.0(7) & 13.8(3)  & \\
$v$=0 J=14-13 & 249435.519 & 249435.5(3) & 20.6(5) & 13.4(4)  & \\
$v$=0 J=15-14 & 267242.336 & 267242.0(4) & 23.2(6) & 14.6(5)  & \\
$v$=0 J=16-15 & 285047.083 & 285046.8(5) & 24.9(10)& 13.7(6)  & \\
$v$=0 J=17-16 & 302849.622 & 302850.3(7) & 22.8(8) & 13.9(7)  & \\
$v$=0 J=18-17 & 320649.814 & 320649.1(5) & 20.0(7) & 13.9(6)  & \\
$v$=0 J=19-18 & 338447.522 & 338447.5(5) & 11.7(5) & 14.2(5)  & \\
$v$=1 J=9-8   & 159602.038 & 159600.6(15)& 0.12(3) & 8.1(10)  & $!$ \\
$v$=1 J=14-13 & 248232.767 & 248232.9(5) & 0.27(4) & 6.1(6)   & \\
$v$=1 J=16-15 & 283672.487 & 283673.7(8) & 0.38(2) & 5.6(6)   & \\
$v$=1 J=17-16 & 301389.100 & 301390.1(6) & 0.46(3) & 6.4(7)   & \\
$v$=1 J=18-17 & 319103.365 & 319105.5(20)& 0.50(6) & 7.5(10)  & \\
$v$=1 J=19-18 & 336815.143 & 336816.8(15)& 0.40(4) & 10.1(10) & \\
$v$=2 J=12-11 & 211754.292 & 211755.1(8) & 0.07(2) & 4.4(7)   & $!$ \\
$v$=2 J=13-12 & 229392.993 & 229392.4(8) & 0.05(1) & 1.4(3)   & $!$ \\
$v$=2 J=16-15 & 282297.754 & 282298.9(15)& 0.08(1) & 2.6(8)   & \$ \\
$v$=2 J=17-16 & 299928.433 & 299928.8(10)& 0.07(2) & 2.2(5)   & \\
$v$=2 J=18-17 & 317556.761 & 317556.0(10)& 0.17(3) & 7.6(10)  & \\
\hline
\multicolumn{6}{c}{$^{30}$SiS} \\
$v$=0 J=5-4   &  87550.623 & 87550.5(3)  & 2.5(1)  & 14.3(5)  & \\
$v$=0 J=6-5   & 105059.283 & 105059.2(2) & 4.9(1)  & 14.2(3)  & \\
$v$=0 J=8-7   & 140074.069 & 140074.1(4) & 7.7(2)  & 14.7(5)  & \\
$v$=0 J=9-8   & 157579.931 & 157579.8(3) & 12.2(3) & 14.1(3)  & \\
$v$=0 J=10-9  & 175084.593 & 175084.1(3) & 7.4(2)  & 14.3(3)  & \\
$v$=0 J=12-11 & 210089.787 & 210089.6(3) & 14.4(4) & 14.2(3)  & \\
$v$=0 J=13-12 & 227590.052 & 227589.9(3) & 15.7(5) & 13.6(5)  & \\
$v$=0 J=14-13 & 245088.585 & 245088.7(3) & 14.3(5) & 13.9(3)  & \\
$v$=0 J=15-14 & 262585.253 & 262585.0(3) & 13.5(6) & 14.3(3)  & \\
$v$=0 J=16-15 & 280079.923 & 280080.8(6) & 16.2(8) & 14.0(5)  & \\
$v$=0 J=17-16 & 297572.461 & 297572.1(5) & 13.2(6) & 14.0(4)  & \\
$v$=0 J=18-17 & 315062.733 & 315062.1(5) & 11.8(5) & 13.5(5)  & \\
$v$=0 J=19-18 & 332550.608 & 332549.9(7) & 10.6(6) & 14.2(5)  & \\
$v$=0 J=20-19 & 350035.950 & 350035.7(4) & 8.1(3)  & 13.8(5)  & \\
$v$=1 J=8-7   & 139404.721 & 139404.4(8) & 0.05(1) & 5.1(8)   & \\
$v$=1 J=13-12 & 226502.331 & 226502.3(3) & 0.20(2) & 5.4(6)   & \\
$v$=1 J=14-13 & 243917.185 & 243916.7(5) & 0.24(2) & 6.6(5)   & \\
$v$=1 J=16-15 & 278741.159 & 278741.3(10)& 0.42(5) & 8.8(10)  & \$ \\
$v$=1 J=17-16 & 296150.011 & 296149.2(8) & 0.20(2) & 3.2(7)   & \\
$v$=1 J=19-18 & 330960.781 & 330960.3(6) & 0.31(3) & 3.8(8)   & \\
$v$=2 J=12-11 & 208081.602 & 208083.1(10)& 0.06(1) & 6.4(8)   & \\
$v$=2 J=14-13 & 242745.671 & 242745.2(5) & 0.09(1) & 3.2(5)   & \\
$v$=2 J=17-16 & 294727.422 & 294727.9(7) & 0.09(1) & 3.9(6)   & \\
$v$=3 J=13-12 & 224326.538 & 224327.2(5) & 0.03(1) & 3.2(4)   & $!$ \\
$v$=3 J=14-13 & 241574.006 & 241573.7(6) & 0.05(1) & 4.4(5)   & $!$ \\
\hline
              &            &             &         &          & \\
              &            &             &         &          & \\
              &            &             &         &          & \\
              &            &             &         &          & \\
              &            &             &         &          & \\
              &            &             &         &          & \\
              &            &             &         &          & \\
              &            &             &         &          & \\
              &            &             &         &          & \\
              &            &             &         &          & \\
              &            &             &         &          & \\
              &            &             &         &          & \\
              &            &             &         &          & \\
              &            &             &         &          & \\
              &            &             &         &          & \\
\multicolumn{6}{c}{Si$^{34}$S} \\
$v$=0 J=5-4   & 88285.834  & 88285.6(3)  & 2.71(8) & 14.9(4)  & \$ \\
$v$=0 J=6-5   & 105941.510 & 105941.5(2) & 4.7(1)  & 14.2(4)  & \\
$v$=0 J=8-7   & 141250.286 & 141250.1(3) & 9.7(2)  & 13.9(4)  & \\
$v$=0 J=9-8   & 158903.116 & 158903.0(3) & 16.4(4) & 14.0(4)  & \\
$v$=0 J=10-9  & 176554.726 & 176554.5(3) & 14.5(3) & 13.9(5)  & \\
$v$=0 J=12-11 & 211853.744 & 211853.5(3) & 15.4(3) & 14.1(4)  & \\
$v$=0 J=13-12 & 229500.881 & 229501.0(4) & 15.5(5) & 13.9(7)  & \$ \\
$v$=0 J=14-13 & 247146.257 & 247146.1(3) & 22.2(8) & 14.2(4)  & \\
$v$=0 J=15-14 & 264789.734 & 264789.9(3) & 19.5(5) & 13.8(4)  & \\
$v$=0 J=16-15 & 282431.179 & 282432.1(7) & 22.2(6) & 13.3(5)  & \\
$v$=0 J=17-16 & 300070.455 & 300071.1(7) & 20.1(6) & 13.3(5)  & \\
$v$=0 J=18-17 & 317707.428 & 317707.1(6) & 18.5(6) & 14.0(5)  & \\
$v$=0 J=19-18 & 335341.960 & 335341.8(6) & 12.7(4) & 14.1(5)  & \\
$v$=0 J=20-19 & 352973.918 & 352975.1(8) & 13.3(6) & 13.1(7)  & \\
$v$=1 J=9-8   & 158140.565 & 158140.3(7) & 0.11(2) & 5.6(6)   & \\
$v$=1 J=12-11 & 210836.992 & 210836.6(8) & 0.11(2) & 6.8(6)   & \\
$v$=1 J=14-13 & 245960.030 & 245960.3(4) & 0.13(2) & 5.9(6)   & \\
$v$=1 J=15-14 & 263518.767 & 263518.1(6) & 0.51(4) & 10.8(8)  & \\
$v$=1 J=16-15 & 281075.470 & 281076.4(8) & 0.34(3) & 7.2(10)  & \\
$v$=1 J=17-16 & 298630.002 & 298631.0(15)& 0.49(6) & 6.8(15)  & \$ \\
$v$=1 J=18-17 & 316182.227 & 316184.0(15)& 0.55(6) & 8.6(15)  & \$ \\
$v$=1 J=19-18 & 333732.011 & 333732.2(8) & 0.20(3) & 5.7(8)   & \\
$v$=1 J=20-19 & 351279.217 & 351278.8(6) & 0.35(3) & 9.6(6)   & \\
$v$=2 J=18-17 & 314656.884 & 314657.6(10)& 0.16(2) & 6.3(8)   & \\
$v$=2 J=19-18 & 332121.910 & 332121.2(10)& 0.07(2) & 3.7(10)  & $!$ \\
$v$=3 J=14-13 & 243587.208 & 243586.9(10)& 0.05(1) & 3.0(7)   & $!$ \\
\hline
\multicolumn{6}{c}{Si$^{33}$S} \\
$v$=0 J=5-4   &  89489.232 & 89489.2(3)  & 0.75(5) & 14.1(5)  & \$ \\
$v$=0 J=8-7   & 143175.531 & 143175.5(5) & 2.4(1)  & 14.5     & $\|$ \$ \\
$v$=0 J=9-8   & 161068.923 & 161069.0(3) & 2.48(7) & 14.0(3)  & \\
$v$=0 J=10-9  & 178961.062 & 178961.0(5) & 3.36(9) & 14.1(4)  & \$ \\
$v$=0 J=11-10 & 196851.808 & 196851.9(5) & 2.05(8) & 14.6(5)  & \\
$v$=0 J=12-11 & 214741.024 & 214741.0(4) & 2.27(7) & 14.5(4)  & \\
$v$=0 J=13-12 & 232628.569 & 232628.4(3) & 3.5(1)  & 14.1(3)  & \\
$v$=0 J=14-13 & 250514.304 & 250514.4(4) & 1.71(7) & 14.1(4)  & \\
$v$=0 J=15-14 & 268398.090 & 268398.7(6) & 5.0(4)  & 14.5(4)  & \\
$v$=0 J=16-15 & 286279.788 & 276279.3(8) & 3.1(2)  & 13.8(8)  & \$ \\
$v$=0 J=17-16 & 304159.258 & 304159.1(4) & 4.8(2)  & 13.2(6)  & \\
$v$=0 J=18-17 & 322036.362 & 322036.6(8) & 3.8(2)  & 13.9(7)  & \\
$v$=0 J=19-18 & 339910.961 & 339910.8(7) & 2.52(8) & 13.7(6)  & \\
$v$=1 J=17-16 & 302689.236 & 302690.2(10)& 0.18(3) & 4.6(8)   & \$ \\
$v$=1 J=18-17 & 320479.853 & 320479.5(10)& 0.05(1) & 1.9(8)   & $!$ \\
\end{longtable}
}

\onltab{4}{
\begin{table*} \caption{NaCl line parameters in IRC
+10216} \label{table-nacl-line-param} \centering
\begin{tabular}{crrcll}
\hline \hline
\multicolumn{1}{c}{Transition} & \multicolumn{1}{c}{$\nu_{\rm rest}$} & \multicolumn{1}{c}{$\nu_{\rm obs}$} & \multicolumn{1}{c}{$\int$T$_{A}^*d$v} & \multicolumn{1}{c}{v$_{\rm exp}$} & \multicolumn{1}{c}{Notes} \\
                               & \multicolumn{1}{c}{(MHz)}            & \multicolumn{1}{c}{(MHz)}           & \multicolumn{1}{c}{K km s$^{-1}$}     & \multicolumn{1}{c}{km s$^{-1}$}   & \\
\hline
\multicolumn{6}{c}{Na$^{35}$Cl} \\
$v$=0 J=7-6   &  91169.875 & 91170.0(3)  & 1.12(5) & 15.3(5)  & \\
$v$=0 J=8-7   & 104189.660 & 104189.6(4) & 1.12(7) & 13.7(7)  & \\
$v$=0 J=10-9  & 130223.627 & 130223.5(3) & 1.78(8) & 13.8(6)  & \\
$v$=0 J=11-10 & 143237.360 & 143237.4(3) & 1.95(8) & 13.1(6)  & \\
$v$=0 J=12-11 & 156248.627 & 156248.6(4) & 2.26(10)& 13.1(7)  & \\
$v$=0 J=13-12 & 169257.205 & 169260.1(20)& 1.24(20)& 10.8(15) & $!$ \\
$v$=0 J=16-15 & 208264.556 & 208264.6(3) & 2.18(10)& 14.3(8)  & \$ \\
$v$=0 J=17-16 & 221260.132 & 221256.0(3) & 2.44(8) & 14.2(4)  & \\
$v$=0 J=18-17 & 234251.898 & 234251.2(4) & 2.14(8) & 14.8(4)  & \\
$v$=0 J=19-18 & 247239.628 & 247239.4(4) & 2.63(6) & 13.3(5)  & \\
$v$=0 J=20-19 & 260223.098 & 260223.8(8) & 4.5(2)  & 14.5(7)  & \\
$v$=0 J=21-20 & 273202.085 & 273204.1(15)& 2.44(10)& 13.2(8)  & \\
$v$=0 J=22-21 & 286176.364 & 286176.2(8) & 2.03(8) & 14.3(7)  & \\
$v$=0 J=23-22 & 299145.710 & 299145.8(6) & 1.90(6) & 15.9(10) & \\
$v$=0 J=24-23 & 312109.900 & 312109.8(6) & 1.39(5) & 12.2(6)  & \\
$v$=0 J=26-25 & 338021.912 & 338021.6(7) & 1.13(4) & 13.2(7)  & \\
$v$=0 J=27-26 & 350969.286 & 350968.9(7) & 1.14(4) & 13.1(6)  & \\

\hline
\multicolumn{6}{c}{Na$^{37}$Cl} \\
$v$=0 J=7-6   & 89220.115  & 89219.9(5)  & 0.39(3) & 15.7(10) & \\
$v$=0 J=8-7   & 101961.552 & 101961.2(4) & 0.41(2) & 13.6(6)  & \\
$v$=0 J=9-8   & 114701.273 & 114701.1(8) & 0.43(5) & 13.0(15) & \$ \\
$v$=0 J=16-15 & 203813.203 & 203812.0(15)& 0.55(7) & 14.0(10) & \\
$v$=0 J=20-19 & 254663.465 & 254662.8(15)& 0.74(10)& 11.9(20) & \$ \\
$v$=0 J=21-20 & 267365.833 & 267365.9(8) & 0.83(7) & 13.3(8)  & \\
$v$=0 J=23-22 & 292756.828 & 292756.1(7) & 0.89(6) & 12.5(7)  & \\
$v$=0 J=24-23 & 305445.026 & 305444.3(8) & 1.16(7) & 12.4(6)  & \\
$v$=0 J=25-24 & 318128.071 & 318127.4(10)& 0.43(5) & 15.0(10) & \\
$v$=0 J=26-25 & 330805.749 & 330805.5(10)& 0.57(5) & 10.9(10) & \\
$v$=0 J=27-26 & 343477.843 & 343478.4(10)& 0.67(5) & 15.2(8)  & \\
\hline
\end{tabular}
\end{table*}
}

\onltab{5}{
\begin{table*} \caption{KCl line parameters in IRC
+10216} \label{table-kcl-line-param} \centering
\begin{tabular}{crrcll}
\hline \hline
\multicolumn{1}{c}{Transition} & \multicolumn{1}{c}{$\nu_{\rm rest}$} & \multicolumn{1}{c}{$\nu_{\rm obs}$} & \multicolumn{1}{c}{$\int$T$_{A}^*d$v} & \multicolumn{1}{c}{v$_{\rm exp}$} & \multicolumn{1}{c}{Notes} \\
                               & \multicolumn{1}{c}{(MHz)}            & \multicolumn{1}{c}{(MHz)}           & \multicolumn{1}{c}{K km s$^{-1}$}     & \multicolumn{1}{c}{km s$^{-1}$}   & \\
\hline
\multicolumn{6}{c}{K$^{35}$Cl} \\
$v$=0 J=11-10 &  84562.590 &  84562.6(8) & 0.14(3) & 13.5(10) & \\
$v$=0 J=12-11 &  92246.501 &  92246.6(6) & 0.20(2) & 14.5     & $\|$ \\
$v$=0 J=13-12 &  99929.473 &  99928.8(10)& 0.21(3) & 14.0(8)  & \\
$v$=0 J=14-13 & 107611.429 & 107612.6(10)& 0.32(5) & 12.2(10) & \$ \\
$v$=0 J=17-16 & 130650.412 & 130650.3(8) & 0.35(6) & 14.2(8)  & \\
$v$=0 J=19-18 & 146003.215 & 146001.9(10)& 0.37(6) & 11.3(10) & \\
$v$=0 J=20-19 & 153677.427 & 153677.8(15)& 0.34(8) & 15.5(15) & $!$ \\
$v$=0 J=23-22 & 176690.360 & 176692.2(15)& 0.62(10)& 16.0(15) & \\
$v$=0 J=27-26 & 207348.712 & 207347.8(10)& 0.48(5) & 11.3(10) & \\
$v$=0 J=28-27 & 215008.213 & 215008.8(6) & 0.40(3) & 14.7(5)  & \\
$v$=0 J=29-28 & 222665.524 & 222665.3(5) & 0.68(3) & 12.1(6)  & \\
$v$=0 J=30-29 & 230320.564 & 230320.5(8) & 0.60(3) & 13.2(6)  & \\
$v$=0 J=31-30 & 237973.257 & 237972.0(10)& 0.46(4) & 13.6(8)  & \\
$v$=0 J=33-32 & 253271.284 & 253271.1(5) & 0.90(4) & 14.0(6)  & \\
$v$=0 J=34-33 & 260916.462 & 260916.2(10)& 0.58(4) & 11.8(8)  & \\
$v$=0 J=36-35 & 276198.754 & 276198.7(15)& 0.54(6) & 12.3(15) & \\
$v$=0 J=37-36 & 283835.711 & 283838.0(20)& 0.80(3) & 13.2(5)  & \\
$v$=0 J=39-38 & 299100.855 & 299105.6(20)& 0.71(8) & 16.7(20) & \\
$v$=0 J=40-39 & 306728.885 & 306727.9(10)& 0.85(7) & 15.7(10) & \\
$v$=0 J=41-40 & 314353.781 & 314354.4(10)& 0.66(5) & 15.3(10) & \\
$v$=0 J=42-41 & 321975.467 & 321975.4(15)& 0.34(5) & 7.1(15)  & \\
$v$=0 J=43-42 & 329593.862 & 329594.5(10)& 0.62(6) & 10.5(10) & \\
$v$=0 J=44-43 & 337208.889 & 337209.4(10)& 0.40(3) & 13.6(10) & \\
$v$=0 J=45-44 & 344820.468 & 344818.4(20)& 0.66(9) & 17.5(20) & \\
\hline
\multicolumn{6}{c}{K$^{37}$Cl} \\
$v$=0 J=11-10 &  82159.153 &  82159.0(10)& 0.08(2) & 14.5     & $\|$ $!$ \\
$v$=0 J=12-11 &  89624.771 &  89625.0(10)& 0.08(2) & 14.5     & $\|$ \$ $!$ \\
$v$=0 J=13-12 &  97089.504 &  97089.0(7) & 0.08(2) & 13.3(8)  & \\
$v$=0 J=14-13 & 104553.276 & 104553.3(15)& 0.11(3) & 14.5     & $\|$ \$ $!$ \\
$v$=0 J=15-14 & 112016.016 & 112017.1(15)& 0.06(2) & 14.5     & $\|$ \$ $!$ \\
$v$=0 J=28-27 & 208903.957 & 208903.2(15)& 0.18(2) & 14.5     & $\|$ $!$ \\
$v$=0 J=30-29 & 223782.837 & 223782.8(10)& 0.19(2) & 10.4(10) & \\
$v$=0 J=33-32 & 246084.350 & 246084.2(10)& 0.12(2) & 11.0(10) & $!$ \\
$v$=0 J=35-34 & 260939.958 & 260942.4(20)& 0.42(8) & 12.0(20) & \\
$v$=0 J=36-35 & 268363.919 & 268365.2(20)& 0.27(4) & 14.5     & $\|$ \$ \\
$v$=0 J=39-38 & 290619.543 & 290619.2(20)& 0.42(6) & 11.6(20) & \\
$v$=0 J=40-39 & 298032.418 & 298034.7(20)& 0.19(3) & 7.1(15)  & \\
$v$=0 J=43-42 & 320253.005 & 320251.0(20)& 0.22(3) & 12.0(20) & \\
$v$=0 J=45-44 & 335050.955 & 335051.0(20)& 0.14(2) & 6.4(20)  & $!$ \\
\hline
\end{tabular}
\end{table*}
}

\onltab{6}{
\begin{table*} \caption{AlCl line parameters in IRC
+10216} \label{table-alcl-line-param} \centering
\begin{tabular}{crrcll}
\hline \hline
\multicolumn{1}{c}{Transition} & \multicolumn{1}{c}{$\nu_{\rm rest}$} & \multicolumn{1}{c}{$\nu_{\rm obs}$} & \multicolumn{1}{c}{$\int$T$_{A}^*d$v} & \multicolumn{1}{c}{v$_{\rm exp}$} & \multicolumn{1}{c}{Notes} \\
                               & \multicolumn{1}{c}{(MHz)}            & \multicolumn{1}{c}{(MHz)}           & \multicolumn{1}{c}{K km s$^{-1}$}     & \multicolumn{1}{c}{km s$^{-1}$}   & \\
\hline
\multicolumn{6}{c}{Al$^{35}$Cl} \\
$v$=0 J=6-5   &  87458.226 &  87458.1(3) & 0.53(4) & 14.5(8)  & \$ \\
$v$=0 J=7-6   & 102031.870 & 102031.6(2) & 0.90(4) & 14.9(6)  & \\
$v$=0 J=10-9  & 145744.527 & 145744.6(5) & 2.85(10)& 14.2(7)  & \$ \\
$v$=0 J=11-10 & 160312.055 & 160312.0(2) & 3.31(7) & 14.0(4)  & \\
$v$=0 J=12-11 & 174877.605 & 174877.4(4) & 2.21(10)& 13.9(8)  & \$ \\
$v$=0 J=14-13 & 204002.051 & 204002.0(4) & 3.13(5) & 14.0(6)  & \\
$v$=0 J=15-14 & 218560.586 & 218560.5(5) & 3.20(6) & 13.8(5)  & \\
$v$=0 J=16-15 & 233116.423 & 233115.7(5) & 5.2(1)  & 14.2(4)  & \\
$v$=0 J=17-16 & 247669.382 & 247669.4(5) & 6.8(2)  & 14.2(5)  & \$ \\
$v$=0 J=18-17 & 262219.283 & 262219.0(8) & 6.6(2)  & 14.8(8)  & \$ \\
$v$=0 J=19-18 & 276765.946 & 276767.1(6) & 5.9(1)  & 14.1(5)  & \\
$v$=0 J=20-19 & 291309.191 & 291309.7(6) & 6.6(1)  & 14.3(5)  & \\
$v$=0 J=21-20 & 305848.837 & 305848.9(5) & 6.4(1)  & 14.4(5)  & \\
$v$=0 J=22-21 & 320384.706 & 320384.1(6) & 5.8(1)  & 13.2(7)  & \\
$v$=0 J=23-22 & 334916.617 & 334916.7(5) & 4.45(10)& 13.7(6)  & \\
$v$=0 J=24-23 & 349444.289 & 349443.3(10)& 7.4(2)  & 14.5     & $\|$ \$ \\
\hline
\multicolumn{6}{c}{Al$^{37}$Cl} \\
$v$=0 J=6-5   &  85403.976 &  85403.4(8) & 0.11(2) & 14.5     & $\|$ \\
$v$=0 J=7-6   &  99635.372 &  99635.5(4) & 0.35(3) & 14.6(7)  & \\
$v$=0 J=8-7   & 113865.568 & 113865.4(7) & 0.29(3) & 13.3(10) & \\
$v$=0 J=10-9  & 142321.673 & 142322.5(10)& 0.84(8) & 10.9(10) & \\
$v$=0 J=11-10 & 156547.240 & 156547.1(10)& 1.77(15)& 14.5(15) & \$ \\
$v$=0 J=15-14 & 213428.929 & 213428.6(4) & 2.19(5) & 14.0(5)  & \\
$v$=0 J=16-15 & 227643.350 & 227643.3(4) & 2.40(4) & 14.3(4)  & \\
$v$=0 J=17-16 & 241855.027 & 241855.2(3) & 2.26(4) & 13.8(3)  & \\
$v$=0 J=18-17 & 256063.788 & 256063.8(4) & 1.52(3) & 13.8(4)  & \\
$v$=0 J=19-18 & 270269.462 & 270269.2(5) & 2.53(3) & 14.2(5)  & \\
$v$=0 J=20-19 & 284471.877 & 284473.1(8) & 2.66(3) & 14.0(5)  & \\
$v$=0 J=21-20 & 298670.862 & 298672.0(8) & 1.79(2) & 13.9(5)  & \\
$v$=0 J=22-21 & 312866.245 & 312865.3(10)& 1.79(3) & 14.5     & $\|$ \$ \\
$v$=0 J=23-22 & 327057.854 & 327057.1(8) & 2.46(4) & 13.8(8)  & \\
$v$=0 J=24-23 & 341245.517 & 341245.2(6) & 1.66(3) & 13.2(6)  & \\
$v$=0 J=25-24 & 355429.063 & 355430.9(10)& 1.68(4) & 14.2(8)  & \\
\hline
\end{tabular}
\end{table*}
}

\onltab{7}{
\begin{table*} \caption{AlF line parameters in IRC
+10216} \label{table-alf-line-param} \centering
\begin{tabular}{crrcll}
\hline \hline
\multicolumn{1}{c}{Transition} & \multicolumn{1}{c}{$\nu_{\rm rest}$} & \multicolumn{1}{c}{$\nu_{\rm obs}$} & \multicolumn{1}{c}{$\int$T$_{A}^*d$v} & \multicolumn{1}{c}{v$_{\rm exp}$} & \multicolumn{1}{c}{Notes} \\
                               & \multicolumn{1}{c}{(MHz)}            & \multicolumn{1}{c}{(MHz)}           & \multicolumn{1}{c}{K km s$^{-1}$}     & \multicolumn{1}{c}{km s$^{-1}$}   & \\
\hline
\multicolumn{6}{c}{AlF} \\
$v$=0 J=3-2   &  98926.768 &  98926.8(5) & 0.93(8) & 14.1(7)  & \$ \\
$v$=0 J=4-3   & 131898.841 & 131899.1(15)& 1.95(15)& 17.0(15) & \\
$v$=0 J=5-4   & 164867.899 & 164867.4(8) & 2.8(1)  & 12.8(8)  & \\
$v$=0 J=6-5   & 197833.190 & 197833.1(3) & 2.6(1)  & 14.4(4)  & \\
$v$=0 J=7-6   & 230793.958 & 230793.6(3) & 3.5(1)  & 14.3(5)  & \\
$v$=0 J=8-7   & 263749.452 & 263749.3(4) & 5.2(1)  & 14.5(4)  & \\
$v$=0 J=9-8   & 296698.916 & 296699.0(3) & 4.7(1)  & 13.8(4)  & \\
$v$=0 J=10-9  & 329641.598 & 329640.0(8) & 5.1(1)  & 14.3(5)  & \\
\hline
\end{tabular}
\end{table*}
}

\onllongtab{8}{
\begin{longtable}{ccrcll}
\caption{NaCN line parameters in IRC +10216} \label{table-nacn-line-param}\\
\hline \hline
\multicolumn{1}{c}{Transition} & \multicolumn{1}{c}{$\nu_{\rm rest}$} & \multicolumn{1}{c}{$\nu_{\rm obs}$} & \multicolumn{1}{c}{$\int$T$_{A}^*d$v} & \multicolumn{1}{c}{v$_{\rm exp}$} & \multicolumn{1}{c}{Notes} \\
                               & \multicolumn{1}{c}{(MHz)}            & \multicolumn{1}{c}{(MHz)}           & \multicolumn{1}{c}{K km s$^{-1}$}     & \multicolumn{1}{c}{km s$^{-1}$}   & \\
\hline
\endfirsthead
\caption{Continued.} \\
\hline
\multicolumn{1}{c}{Transition} & \multicolumn{1}{c}{$\nu_{\rm rest}$} & \multicolumn{1}{c}{$\nu_{\rm obs}$} & \multicolumn{1}{c}{$\int$T$_{A}^*d$v} & \multicolumn{1}{c}{v$_{\rm exp}$} & \multicolumn{1}{c}{Notes} \\
                               & \multicolumn{1}{c}{(MHz)}            & \multicolumn{1}{c}{(MHz)}           & \multicolumn{1}{c}{K km s$^{-1}$}     & \multicolumn{1}{c}{km s$^{-1}$}   & \\
\hline
\endhead
\hline
\endfoot
\hline
\endlastfoot
\multicolumn{6}{c}{NaCN} \\
5$_{1,4}$-4$_{1,3}$    &  80846.901 &  80846.9(3) & 0.79(5) & 13.7(5)  & \\
6$_{1,6}$-5$_{1,5}$    &  90394.384 &  90394.1(3) & 0.93(5) & 14.8(4)  & \\
6$_{0,6}$-5$_{0,5}$    &  93206.092 &  93206.0(3) & 1.10(7) & 14.2(4)  & \\
6$_{5,2}$-5$_{5,1}$ + 6$_{5,1}$-5$_{5,0}$ &  93637.583 + 93637.584 &  93637.1(10)& 0.14(2) & 14.4(10) & $\pounds$ \\
6$_{2,5}$-5$_{2,4}$    &  93712.543 &  93712.4(10)& 0.65(15)& 14.5     & $\|$ \$ \\
6$_{4,3}$-5$_{4,2}$ + 6$_{4,2}$-5$_{4,1}$ &  93738.970 + 93739.025 &  93738.5(4) & 0.55(4) & 12.7(8)  & $\pounds$ \\
6$_{3,4}$-5$_{3,3}$    &  93838.443 &  93838.5(7) & 0.46(6) & 13.8(6)  & \$ \\
6$_{3,3}$-5$_{3,2}$    &  93848.623 &  93848.8(7) & 0.44(6) & 14.3(6)  & \$ \\
6$_{2,4}$-5$_{2,3}$    &  94334.803 &  94334.9(3) & 1.02(6) & 14.6(4)  & \\
6$_{1,5}$-5$_{1,4}$    &  96959.808 &  96959.9(3) & 1.00(6) & 14.0(4)  & \\
7$_{1,7}$-6$_{1,6}$    & 105393.275 & 105393.0(3) & 1.24(8) & 14.7(5)  & \$ \\
7$_{0,7}$-6$_{0,6}$    & 108471.987 & 108472.0(2) & 1.27(7) & 14.7(4)  & \\
7$_{5,3}$-6$_{5,2}$ + 7$_{5,2}$-6$_{5,1}$ & 109250.132 + 109250.133 &  109249.6(6) & 0.34(2) & 14.2(8)  & $\pounds$ \\
7$_{2,6}$-6$_{2,5}$    & 109281.478 & 109281.5(8) & 0.86(8) & 14.0(6)  & \$ \\
7$_{4,4}$-6$_{4,3}$ + 7$_{4,3}$-6$_{4,2}$ & 109375.276 + 109375.458 &  109375.2(2) & 0.66(4) & 15.2(5)  & $\pounds$ \\
7$_{3,5}$-6$_{3,4}$    & 109501.960 & 109501.9(6) & 0.62(5) & 14.5     & $\|$ \$ \\
7$_{3,4}$-6$_{3,3}$    & 109524.843 & 109525.5(6) & 0.52(4) & 14.5     & $\|$ \$ \\
7$_{2,5}$-6$_{2,4}$    & 110269.912 & 110270.0(3) & 1.19(6) & 14.9(5)  & \$ \\
7$_{1,6}$-6$_{1,5}$    & 113040.219 & 113040.1(6) & 0.96(6) & 14.3(6)  & \$ \\
8$_{1,7}$-7$_{1,6}$    & 129081.302 & 129081.6(8) & 1.72(10)& 13.8(10) & \\
9$_{1,9}$-8$_{1,8}$    & 135303.358 & 135302.4(10)& 1.22(12)& 15.9(15) & \\
9$_{0,9}$-8$_{0,8}$    & 138652.095 & 138652.0(5) & 1.33(6) & 13.9(5)  & \\
9$_{4,6}$-8$_{4,5}$ + 9$_{4,5}$-8$_{4,4}$ & 140666.698 + 140667.897 & 140666.3(12) & 0.67(8) & 13.4(10)  & $\pounds$ \\
9$_{2,7}$-8$_{2,6}$    & 142410.522 & 142410.4(5) & 0.86(5) & 14.3(4)  & \\
9$_{1,8}$-8$_{1,7}$    & 145075.566 & 145075.3(5) & 1.22(6) & 13.8(6)  & \$ \\
10$_{0,10}$-9$_{0,9}$  & 153557.927 & 153557.0(10)& 1.13(7) & 15.2(10) & \\
10$_{2,9}$-9$_{2,8}$   & 155837.923 & 155838.1(8) & 1.06(8) & 13.4(10) & \\
10$_{4,7}$-9$_{4,6}$ + 10$_{4,6}$-9$_{4,5}$ & 156323.334 + 156325.930 & 156324.5(10) & 1.46(10) & 13.7(10)  & $\pounds$ \\
10$_{3,8}$-9$_{3,7}$   & 156541.550 & 156540.9(15)& 0.66(9) & 14.5(10) & \$ \\
10$_{3,7}$-9$_{3,6}$   & 156684.358 & 156684.7(8) & 0.86(6) & 11.0(6)  & \\
10$_{2,8}$-9$_{2,7}$   & 158616.767 & 158617.2(8) & 1.07(8) & 14.8(8)  & \\
10$_{1,9}$-9$_{1,8}$   & 161014.794 & 161014.8(5) & 1.13(5) & 14.1(5)  & \\
11$_{5,7}$-10$_{5,6}$ + 11$_{5,6}$-10$_{5,5}$ & 171733.484 + 171733.540 & 171735.4(10) & 0.89(12) & 9.1(12)  & $\pounds$ \\
11$_{2,9}$-10$_{2,8}$   & 174904.123 & 174903.3(5) & 0.48(3) & 13.5(4)  & \\
11$_{1,10}$-10$_{1,9}$  & 176889.996 & 176889.1(8) & 0.64(6) & 13.4(8)  & \\
12$_{1,12}$-11$_{1,11}$ & 179921.906 & 179920.8(12)& 0.96(10)& 12.6(15) & \\
13$_{0,13}$-12$_{0,12}$ & 197611.467 & 197611.8(6) & 0.53(3) & 13.0(7)  & \\
13$_{2,12}$-12$_{2,11}$ & 202103.717 & 202106.9(20)& 1.09(5) & 16.0(20) & \\
13$_{6,8}$-12$_{6,7}$ + 13$_{6,7}$-12$_{6,6}$ & 202688.585 + 202688.587 & 202685.9(20) & 0.65(5) & 14.0(15) & $\pounds$ \\
13$_{5,9}$-12$_{5,8}$ + 13$_{5,8}$-12$_{5,7}$ & 203000.858 + 203001.131 & 203002.8(10) & 0.91(6) & 10.9(15) & $\pounds$ \\
13$_{3,11}$-12$_{3,10}$ & 203627.152 & 203627.8(10)& 0.82(6) & 15.2(10) & \\
13$_{3,10}$-12$_{3,9}$  & 204158.738 & 204158.6(10)& 0.45(4) & 11.6(10) & \\
13$_{2,11}$-12$_{2,10}$ & 207658.257 & 207657.0(10)& 0.82(6) & 13.6(10) & \\
13$_{1,12}$-12$_{1,11}$ & 208408.609 & 208408.6(6) & 0.76(5) & 13.3(6)  & \\
14$_{1,14}$-13$_{1,13}$ & 209495.655 & 209497.4(15)& 0.39(5) & 11.5(15) & \$ \\
14$_{0,14}$-13$_{0,13}$ & 212124.029 & 212123.1(10)& 0.52(5) & 15.4(10) & \\
14$_{3,12}$-13$_{3,11}$ & 219321.048 & 219323.5(10)& 0.44(8) & 13.1(10) & \\
14$_{3,11}$-13$_{3,10}$ & 220086.453 & 220086.2(15)& 0.32(6) & 13.2(10) & \\
14$_{1,13}$-13$_{1,12}$ & 224030.638 & 224030.5(5) & 0.52(4) & 12.4(6)  & \\
14$_{2,12}$-13$_{2,11}$ & 224082.616 & 224082.2(5) & 0.62(5) & 13.2(6)  & \\
15$_{1,15}$-14$_{1,14}$ & 224232.257 & 224232.1(6) & 0.51(4) & 14.1(6)  & \\
15$_{0,15}$-14$_{0,14}$ & 226581.137 & 226580.3(10)& 0.67(4) & 13.3(6)  & \\
15$_{2,14}$-14$_{2,13}$ & 232748.129 & 232747.6(10)& 0.68(6) & 14.1(8)  & \\
15$_{7,9}$-14$_{7,8}$ + 15$_{7,8}$-14$_{7,7}$    & 233510.583 + 233510.583 & 233510.0(10) & 0.30(3) & 13.0(10) & $\pounds$ \\
15$_{6,10}$-14$_{6,9}$ + 15$_{6,9}$-14$_{6,8}$   & 233895.500 + 233895.513 & 233896.2(10) & 0.39(3) & 10.9(10) & $\pounds$ \\
15$_{5,11}$-14$_{5,10}$ + 15$_{5,10}$-14$_{5,9}$ & 234289.734 + 234290.771 & 234290.4(10) & 0.50(3) & 12.6(8)  & $\pounds$ \\
15$_{4,12}$-14$_{4,11}$ & 234738.183 & 234737.5(8) & 0.30(3) & 12.9(8)  & \\
15$_{3,12}$-14$_{3,11}$ & 236078.907 & 236079.6(8) & 0.20(2) & 11.1(10) & \\
15$_{1,14}$-14$_{1,13}$ & 239546.409 & 239546.2(6) & 0.58(4) & 12.5(6)  & \\
15$_{2,13}$-14$_{2,12}$ & 240507.236 & 240506.8(5) & 0.58(4) & 13.4(6)  & \\
16$_{2,15}$-15$_{2,14}$ & 248003.167 & 248002.7(5) & 0.55(4) & 13.3(6)  & \\
16$_{7,10}$-15$_{7,9}$ + 16$_{7,9}$-15$_{7,8}$    & 249079.914 + 249079.915 & 249079.1(8) & 0.49(6) & 14.2(8)  & $\pounds$ \\
16$_{6,11}$-15$_{6,10}$ + 16$_{6,10}$-15$_{6,9}$  & 249502.758 + 249502.785 & 249503.2(8) & 0.64(7) & 12.9(8)  & $\pounds$ \\
16$_{5,12}$-15$_{5,11}$ + 16$_{5,11}$-15$_{5,10}$ & 249943.220 + 249945.104 & 249944.0(6) & 0.69(7) & 12.4(6)  & $\pounds$ \\
17$_{1,17}$-16$_{1,16}$ & 253611.361 & 253610.6(8) & 0.54(6) & 12.8(8)  & \\
17$_{0,17}$-16$_{0,16}$ & 255396.376 & 255397.6(8) & 0.78(7) & 14.0(5)  & \\
17$_{10,7}$-16$_{10,6}$ + 17$_{10,8}$-16$_{10,7}$ & 263154.449 + 263154.449 & 263155.7(20)& 0.36(5) & 18.4(20) & $\pounds$ \$ \\
17$_{2,16}$-16$_{2,15}$ & 263210.770 & 263210.4(8) & 0.38(4) & 14.0(8)  & \\
17$_{9,9}$-16$_{9,8}$ + 17$_{9,8}$-16$_{9,7}$ & 263682.103 + 263682.103  & 263683.6(10)& 0.18(2) & 12.4(10) & $\pounds$ \\
17$_{8,10}$-16$_{8,9}$ + 17$_{8,9}$-16$_{8,8}$ & 264178.070 + 264178.070 & 264177.2(10)& 0.41(3) & 15.8(10) & $\pounds$ \\
17$_{6,12}$-16$_{6,11}$ + 17$_{6,11}$-16$_{6,10}$ & 265112.789 + 265112.843 & 265111.1(15)& 0.66(4) & 15.3(15) & $\pounds$ \\
17$_{5,13}$-16$_{5,12}$ + 17$_{5,12}$-16$_{5,11}$ & 265603.232 + 265606.523 & 265606.4(15)& 0.74(5) & 13.2(15) & $\pounds$ \\
17$_{4,14}$-16$_{4,13}$ & 266168.587 & 266167.4(15)& 0.46(5) & 17.8(15) & \\
17$_{4,13}$-16$_{4,12}$ & 266283.597 & 266281.4(20)& 0.39(6) & 14.5     & $\|$ \$ \\
17$_{3,15}$-16$_{3,14}$ & 266346.964 & 266347.0(10)& 0.61(5) & 14.1(10) & \\
18$_{1,18}$-17$_{1,17}$ & 268257.278 & 268259.9(20)& 0.46(5) & 13.4(15) & \$ \\
17$_{3,14}$-16$_{3,13}$ & 268287.549 & 268288.5(10)& 0.36(4) & 12.3(10) & \\
18$_{0,18}$-17$_{0,17}$ & 269780.806 & 269780.0(8) & 0.48(4) & 11.7(8)  & \\
17$_{1,16}$-16$_{1,15}$ & 270217.365 & 270218.8(10)& 0.58(4) & 12.7(8)  & \\
17$_{2,15}$-16$_{2,14}$ & 273268.589 & 273270.7(15)& 0.50(6) & 10.1(15) & \\
18$_{2,17}$-17$_{2,16}$ & 278369.353 & 278368.5(10)& 0.42(4) & 15.1(10) & \\
18$_{9,10}$-17$_{9,9}$ + 18$_{9,9}$-17$_{9,8}$ & 279177.286 + 279177.286 & 279173.8(20)& 0.26(4) & 10.5(20) & $\pounds$ \\
18$_{8,11}$-17$_{8,10}$ + 18$_{8,10}$-17$_{8,9}$ & 279710.006 + 279710.006 & 279715.1(20)& 0.44(6)& 14.5 & $\pounds$ $\|$ \$ \\
18$_{7,12}$-17$_{7,11}$ + 18$_{7,11}$-17$_{7,10}$ & 280219.750 + 280219.751 & 280220.6(10)& 0.49(5) & 14.6(10) & $\pounds$ \\
18$_{4,14}$-17$_{4,13}$ & 282067.503 & 282070.2(20)& 0.43(5) & 15.2(15) & \\
19$_{1,19}$-18$_{1,18}$ & 282876.644 & 282878.9(20)& 0.47(5) & 13.5(15) & \\
19$_{0,19}$-18$_{0,18}$ & 284161.712 & 284161.3(10)& 0.43(3) & 11.4(8)  & \\
18$_{3,15}$-17$_{3,14}$ & 284514.720 & 284516.3(15)& 0.38(4) & 10.8(10) & \\
18$_{1,17}$-17$_{1,16}$ & 285355.280 & 285353.0(20)& 0.45(5) & 15.3(15) & \\
18$_{2,16}$-17$_{2,15}$ & 289567.202 & 289566.9(10)& 0.48(4) & 13.8(10) & \\
19$_{2,18}$-18$_{2,17}$ & 293477.704 & 293481.9(20)& 0.77(8) & 16.7(15) & \\
19$_{9,11}$-18$_{9,10}$ + 19$_{9,10}$-18$_{9,9}$ & 294669.796 + 294669.796 & 294669.1(20)& 0.33(5) & 13.2(15) & $\pounds$ \\
19$_{7,13}$-18$_{7,12}$ + 19$_{7,12}$-18$_{7,11}$ & 295790.308 + 295790.311 & 295788.6(20)& 0.30(5)& 14.5     & $\pounds$ $\|$ \$ \\
19$_{5,15}$-18$_{5,14}$ + 19$_{5,14}$-18$_{5,13}$ & 296944.126 + 296953.235 & 296948.3(20)& 0.28(4)& 15.0(20) & $\pounds$ \\
20$_{1,20}$-19$_{1,19}$ & 297471.433 & 297470.7(20)& 0.24(4) & 11.0(20) & \\
19$_{4,15}$-18$_{4,14}$ & 297879.787 & 297879.7(15)& 0.20(3) & 12.0(15) & \$ \\
20$_{0,20}$-19$_{0,19}$ & 298544.239 & 298542.5(20)& 0.28(4) & 11.2(15) & \\
19$_{1,18}$-18$_{1,17}$ & 300354.316 & 300354.0(10)& 0.44(4) & 11.7(10) & \\
20$_{8,13}$-19$_{8,12}$ + 20$_{8,12}$-19$_{8,11}$ & 310769.695 + 310769.695 & 310767.2(20)& 0.22(3)& 12.6(20) & $\pounds$ \\
20$_{7,14}$-19$_{7,13}$ + 20$_{7,13}$-19$_{7,12}$ & 311361.332 + 311361.337 & 311361.8(10)& 0.20(2)& 14.2(10) & $\pounds$ \\
20$_{6,15}$-19$_{6,14}$ + 20$_{6,14}$-19$_{6,13}$ & 311961.365 + 311961.708 & 311961.6(10)& 0.35(3)& 15.0(10) & $\pounds$ \\
21$_{1,21}$-20$_{1,20}$ & 312043.593 & 312043.2(8)& 0.33(2) & 12.0(8) & \\
21$_{0,21}$-20$_{0,20}$ & 312931.171 & 312930.1(15)& 0.13(1)& 12.6(15)& \\
20$_{3,18}$-19$_{3,17}$ & 313196.095 & 313198.4(20)& 0.12(2)& 14.5    & $\|$ \\
20$_{4,17}$-19$_{4,16}$ & 313369.371 & 313371.4(15)& 0.22(2)& 16.6(15)& \\
20$_{4,16}$-19$_{4,15}$ & 313724.992 & 313721.0(20)& 0.44(5)& 14.5    & $\|$ \\
20$_{1,19}$-19$_{1,18}$ & 315213.834 & 315212.9(15)& 0.39(4)& 13.9(10)& \\
20$_{3,17}$-19$_{3,16}$ & 317220.575 & 317219.5(10)& 0.38(3)& 12.8(10)& \\
21$_{1,20}$-20$_{1,19}$ & 329937.919 & 329938.4(20)& 0.22(3)& 11.4(20)& \\
23$_{1,23}$-22$_{1,22}$ & 341127.341 & 341128.1(15)& 0.37(3)& 14.8(15)& \\
23$_{0,23}$-22$_{0,22}$ & 341721.282 & 341717.9(20)& 0.30(4)& 12.3(20)& \\
22$_{8,15}$-21$_{8,14}$ + 22$_{8,14}$-21$_{8,13}$ & 341823.295 + 341823.295 & 341823.2(10)& 0.40(4)& 12.7(10) & $\pounds$ \\
22$_{6,17}$-21$_{6,16}$ + 22$_{6,16}$-21$_{6,15}$ & 343211.017 + 343212.019 & 343213.0(20)& 0.42(4)& 15.2(20) & $\pounds$ \\
22$_{3,20}$-21$_{3,19}$ & 344269.527 & 344267.6(20)& 0.19(4)& 11.5(20)& \\
22$_{1,21}$-21$_{1,20}$ & 344535.644 & 344539.8(20)& 0.22(3)& 10.4(20)& \\
22$_{4,19}$-21$_{4,18}$ & 344854.243 & 344857.9(20)& 0.25(4)& 10.2(20)& \\
\end{longtable}
}

\section{Physical model of IRC +10216}
\label{sec-physical-model}

\begin{table}
\caption{IRC +10216's model parameters} \label{table-phys-model}
\centering
\begin{tabular}{l@{\hspace{1.1cm}}r}
\hline \hline
Parameter                                              & Value \\
\hline
Distance ($D$)                                          & 130 pc \\
Stellar radius ($R_*$)                                  & 4 $\times$ 10$^{13}$ cm \\
Stellar effective temperature ($T_*$)                   & 2330 K \\
Stellar luminosity ($L_*$)                              & 8750 L$_{\odot}$ \\
Stellar mass ($M_*$)                                    & 0.8 M$_{\odot}$ \\
End of static atmosphere ($R_0$)                        & 1.2 R$_*$ \\
Dust condensation radius ($R_c$)                        & 5 R$_*$ \\
End of dust acceleration region ($R_w$)                 & 20 R$_*$ \\
Gas expansion velocity ($v_{\rm exp}$) $^a$             & 14.5 km s$^{-1}$ \\
Microturbulence velocity ($\Delta v_{\rm turb}$) $^b$   & 1 km s$^{-1}$ \\
Mass loss rate ($\dot{M}$)                              & 2 $\times$ 10$^{-5}$ M$_{\odot}$ yr$^{-1}$ \\
Gas kinetic temperature ($T_k$) $^c$                    & $T_*$ $\times$ ($r$/$R_*$)$^{-0.55}$ \\
Gas-to-dust mass ratio ($\frac{\rho_g}{\rho_d}$)        & 300 \\
Dust temperature ($T_d$)                                & 800 K $\times$ $(r/R_c)^{-0.375}$ \\
\hline
\end{tabular}
\tablenoteb{\\
$^a$ $v_{\rm exp}$ = 5 km s$^{-1}$ for regions inner
to $R_c$, 11 km s$^{-1}$ for the dust acceleration region between
$R_c$ and $R_w$, and 14.5 km s$^{-1}$ beyond $R_w$. \\
$^b$ $\Delta v_{\rm turb}$ = 5 km s$^{-1}$ $\times$ ($R_*/r$) for
regions inner to $R_c$ and 1 km s$^{-1}$ in the rest of the
envelope. \\
$^c$ $T_k$ $\propto$ $r^{-0.55}$ for regions inner to 75 R$_*$,
$T_k$ $\propto$ $r^{-0.85}$ between 75 and 200 R$_*$, and $T_k$
$\propto$ $r^{-1.40}$ beyond 200 R$_*$.}
\end{table}

The physical model consists of a spherical envelope of gas and
dust expanding around a central AGB star. The parameters adopted
are given in Table~\ref{table-phys-model}. The assumed distance to
IRC +10216 is 130 pc (\cite{men2001} 2001). We consider an AGB
star with an effective temperature $T_*$ of 2330 K (\cite{rid1988}
1988), a radius $R_*$ of 4 $\times$ 10$^{13}$ cm (0.021$''$, which
is consistent with the value of 0.022$''$ derived from infrared
interferometry by \cite{mon2000} 2000). The adopted values for
$T_*$ and $R_*$ imply a stellar luminosity of 8750 L$_{\odot}$,
within the range 5200-13,000 L$_{\odot}$ given by \cite{men2001}
(2001). The adopted stellar mass is 0.8 M$_{\odot}$
(\cite{win1994} 1994; \cite{men2001} 2001).

\begin{figure}
\includegraphics[angle=0,scale=.38]{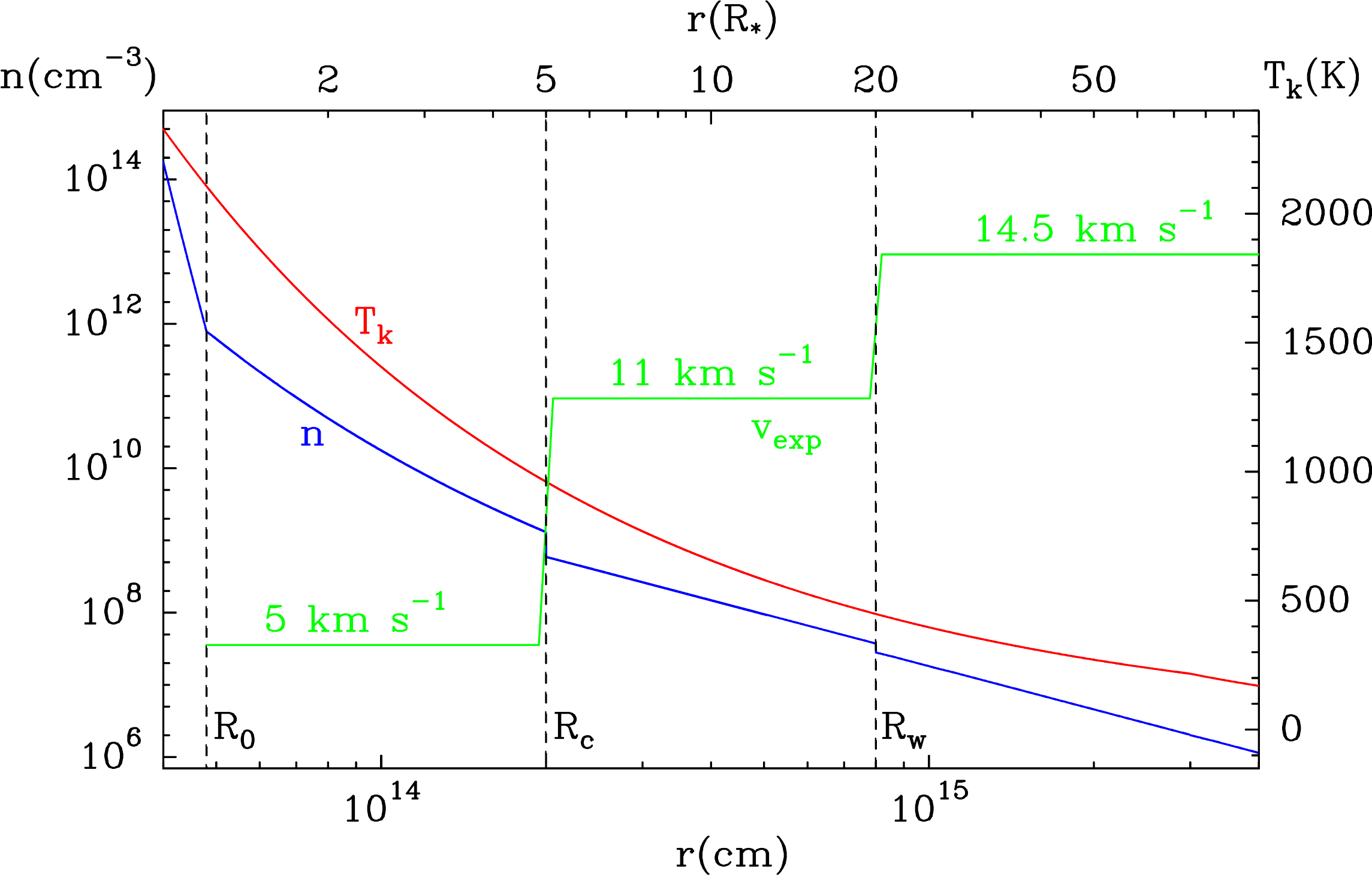}
\caption{Particle density $n$, gas kinetic temperature $T_k$, and
expansion velocity $v_{\rm exp}$ as a function of radius in the
inner layers of IRC +10216.} \label{fig-phys-profile}
\end{figure}

The dust condensation radius $R_c$ is located at 5 R$_*$ and the
end of the dust acceleration region $R_w$ at 20 R$_*$, with the
gas expansion velocity being 5 km s$^{-1}$ for the regions inner
to $R_c$, 11 km s$^{-1}$ in the dust acceleration region, and 14.5
km s$^{-1}$ beyond $R_w$ (see Fig.~\ref{fig-phys-profile}), in
agreement with IR studies of the inner layers of IRC +10216
(\cite{kea1993} 1993; \cite{fon2008} 2008). The adopted
microturbulence velocity $\Delta v_{\rm turb}$ is 5 km s$^{-1}$ at
the stellar photosphere, decreasing down to 1 km s$^{-1}$ at R$_c$
(\cite{kea1988} 1988; see Table~\ref{table-phys-model}). Beyond
R$_c$ we consider a constant microturbulence velocity of 1 km
s$^{-1}$, within the range of values 0.65--1.5 km s$^{-1}$ derived
in the literature (\cite{ski1999} 1999; \cite{deb2012} 2012).

The mass loss rate and gas kinetic temperature in the envelope are
derived from the modeling of the rotational lines $J=1-0$ to
$J=16-15$ of $^{12}$CO and $^{13}$CO observed with the IRAM 30-m
telescope and HIFI (see \cite{deb2012} 2012). We used a non-local
radiative transfer code (\cite{dan2008} 2008), and adopted an
abundance of CO relative to H$_2$ of 6 $\times$ 10$^{-4}$
(\cite{cro1997} 1997; \cite{ski1999} 1999) and an abundance ratio
$^{12}$CO/$^{13}$CO of 45 (\cite{cer2000} 2000). We derive a mass
loss rate of 2 $\times$ 10$^{-5}$ M$_\odot$ yr$^{-1}$, and a gas
kinetic temperature that varies with radius as $r^{-0.55}$ for
regions inner to 75 R$_*$, as $r^{-0.85}$ between 75 and 200
R$_*$, and as $r^{-1.40}$ beyond 200 R$_*$ (see
Fig.~\ref{fig-phys-profile}).

\begin{figure}[b]
\includegraphics[angle=0,scale=.36]{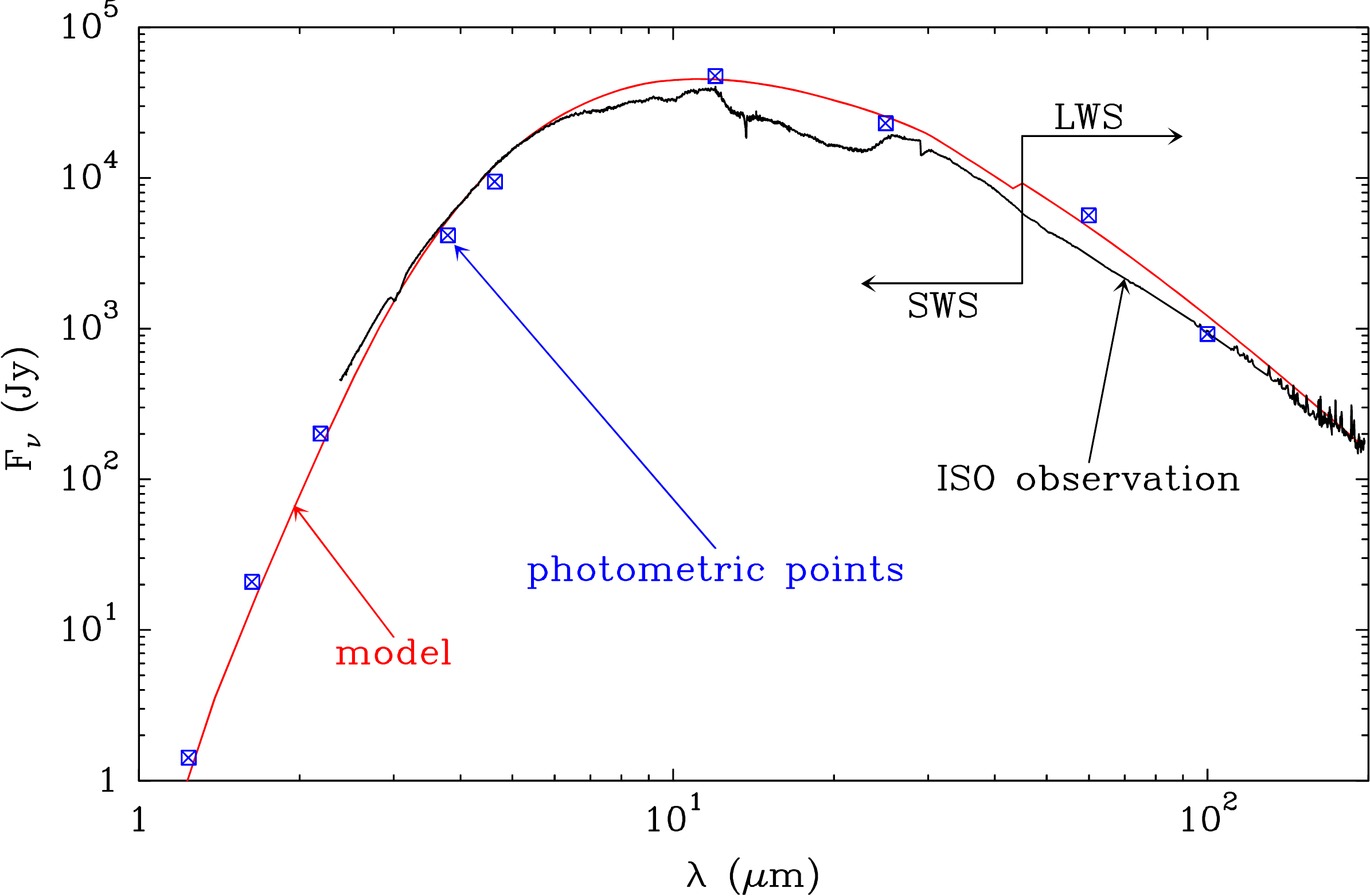}
\caption{Spectral energy distribution of IRC +10216 as observed by
\emph{ISO} (black line; \cite{cer1999} 1999), as given by
photometric measurements (blue crossed boxes; values taken from
\cite{leb1992} (1992) and from the IRAS Point Source Catalog), and
as calculated by the dust model (red line).} \label{fig-sed}
\end{figure}

As concerns dust, we consider spherical grains of amorphous carbon
with a radius of 0.1 $\mu$m, a mass density of 2 g cm$^{-3}$, and
the optical properties of \cite{suh2000} (2000). The dust
continuum is modeled using a ray-tracing code that propagates the
specific intensities along a set of impact parameters which are
subsequently convolved with the telescope beam. The model does not
include scattering, which becomes important only at short
wavelengths, below 5 $\mu$m. From the modeling of the envelope
spectral energy distribution (see Fig.~\ref{fig-sed}) we find a
gas-to-dust mass ratio of 300 and a dust temperature ($T_d$)
radial distribution of the form $T_d$ = 800 K $\times$
$(r/R_c)^{-0.375}$.

The variation of the density of gas particles with radius is
different depending on the region of the envelope. In the static
stellar atmosphere, that extends from the photosphere up to a
radius $R_0$ which we take as 1.2 R$_*$, the density is given by
hydrostatic equilibrium
\begin{equation}
n = n(R_*) \Big( \frac{r}{R_*} \Big)^\alpha \exp \Big\{ - \frac{G
M_* \langle m_{\rm g} \rangle }{k T_* R_* (1-\alpha)} \Big( 1-
\Big( \frac{R_*}{r} \Big)^{(1-\alpha)} \Big) \Big\}
\label{eq-density-hydrostatic-equilibrium}
\end{equation}
where $G$ and $k$ are the gravitational and Boltzmann constants,
respectively, $\alpha$ is the exponent in the kinetic temperature
law ($\alpha$ = 0.55), and $\langle m_{\rm g} \rangle$ is the mean
mass of gas particles (2.3 amu, after considering H$_2$, He, and
CO). In the dynamic atmosphere shock waves induced by the
pulsation of the star extend the circumstellar material. For this
region, that extends from $R_0$ up to $R_c$, we utilize the
equation proposed by \cite{che1992} (1992):
\begin{equation}
n = n(R_0) \exp \Big\{ - \frac{G M_* \langle m_{\rm g} \rangle
(1-\gamma^2)}{k T_* R_*^{\alpha} R_0^{(1-\alpha)} (1-\alpha)}
\Big( 1 - \Big(\frac{r}{R_0} \Big)^{(\alpha-1)} \Big) \Big\}
\label{eq-density-dynamic-atmosphere}
\end{equation}
where $\gamma$ is a dimensionless parameter expressing the ratio
of the shock strength to the local escape velocity ($\gamma$ =
0.89, see \cite{che1992} 1992). Finally, beyond the dust
condensation radius $R_c$, the density at each radius in the
envelope is given by the law of conservation of mass
\begin{equation}
n = \frac{\dot{M}}{(4 \pi r^2 \langle m_{\rm g} \rangle v_{\rm
exp})} \label{eq-density-mass-conservation}
\end{equation}
which results in a density profile varying as $r^{-2}$, with jumps
at $R_c$ and $R_w$, which reflect the jumps in the adopted
expansion velocity profile (see Fig.~\ref{fig-phys-profile}).

The treatment of the density profiles is similar to that adopted
in \cite{agu2006} (2006), except that we have updated some
parameters, such as the stellar mass, and we now do not use
anymore the phenomenological parameter $K_n$ in
Eq.~(\ref{eq-density-dynamic-atmosphere}). The density scale is
fixed by the mass loss rate through
Eq.~(\ref{eq-density-mass-conservation}), and afterwards
Eqs.~(\ref{eq-density-hydrostatic-equilibrium}) and
(\ref{eq-density-dynamic-atmosphere}) determine the density
gradient from $R_c$ down to the photosphere. These two latter
equations involve several uncertain parameters. For example, the
stellar mass is not accurately known although it can be confined
to the range 0.6--1 M$_{\odot}$, based on theoretical arguments
and on the late AGB stage of IRC +10216 (\cite{win1994} 1994).
This relatively small error on the stellar mass translates into an
uncertainty of a factor of 5 for the density at $R_0$. At this
point it is worth to note that densities in the inner layers of
IRC +10216 are not particularly well constrained by observations
and differ by orders of magnitude between different modeling
studies. For example, \cite{wil1998} (1998) adopted, in their
chemical model of the inner wind, a density of 3 $\times$
10$^{11}$ cm$^{-3}$ at 5 R$_*$ rising up to 4 $\times$ 10$^{14}$
cm$^{-3}$ at 1.2 R$_*$, while \cite{sch2006a} (2006a), in their
model aimed at interpreting SiO observations, adopted a value of
about 2 $\times$ 10$^8$ cm$^{-3}$ at 5 R$_*$ rising up to 3
$\times$ 10$^{10}$ cm$^{-3}$ close to the photosphere. In our
model, densities in the inner layers are in between the two above
cases, with a value of 1.8 $\times$ 10$^{14}$ cm$^{-3}$ at $R_*$,
8 $\times$ 10$^{11}$ cm$^{-3}$ at $R_0$, and (0.6--1.3) $\times$
10$^9$ cm$^{-3}$ at $R_c$. The density of particles is therefore
more uncertain in the inner layers than in the outer envelope, and
this uncertainty translates to the molecular abundances derived
(see more details in
Sec.~\ref{subsec-radiative-transfer-results}).

\section{Radiative transfer modeling}

The calculation of the excitation and emergent line profiles for
the different molecules studied here has been done with a
multi--shell radiative transfer program based on the LVG (Large
Velocity Gradient) formalism, which is described in \cite{agu2009}
(2009). Briefly, the spherical circumstellar envelope is divided
into several concentric shells, and statistical equilibrium
equations are solved in each of them. The radiation field
$\overline{J}_{\nu}$ at the frequency ${\nu}$ of each transition,
needed to solve the statistical equilibrium equations, is
evaluated in each shell through an escape probability formalism as
\begin{equation}
\overline{J}_{\nu} = (1 - \beta) S_{\nu} + \beta I_{\nu}^{bg}
\end{equation}
where $S_{\nu}$ is the local source function, $I_{\nu}^{bg}$ is
the specific intensity of the background radiation field, and
$\beta$ is the so-called escape probability, evaluated as
($1-e^{-\tau_\nu}$)/$\tau_\nu$ for an expanding spherical shell
(\cite{cas1970} 1970), with $\tau_\nu$ being the optical depth in
the radial direction within the corresponding shell. The
contributions of both gas and dust are included in the computation
of $S_{\nu}$ and $\tau_\nu$. The background radiation field
arriving at a given shell is composed of the cosmic microwave
background, the stellar component (geometrically diluted and
properly propagated through the inner dusty shells), and the dust
emission arising from the outer shells.

Apart from the radiation at millimeter wavelengths involved in the
observed rotational transitions, infrared radiation plays an
important role for the excitation of some molecules through the
pumping to excited vibrational states. According to the above
description, the source of this infrared radiation in each shell
is on one side local (emission from the warm dust contained in the
shell that contributes to the local source function) and on the
other side external (arising from the surrounding shells
containing warm dust heated by the stellar radiation).

\subsection{Molecular data}
\label{subsec-molecular-data}

The radiative transfer calculations require as input spectroscopic
and collision excitation data. Spectroscopic data has been
obtained from a catalog developed by one of us (J. Cernicharo),
which has been previously used in \cite{cer2000} (2000) and will
be published soon. As concerns collision data, ideally one would
need rate coefficients for inelastic collisions with H$_2$ and He
involving a large enough number of energy levels and covering a
wide enough range of temperatures. Since this is not often the
case, some approximations must be considered. In this study we
have adopted the following rules for all the studied molecules.
Inelastic collisions with both H$_2$ and He (with an abundance of
0.17 relative to H$_2$; \cite{asp2009} 2009) have been considered.
When no rate coefficients for collisions with H$_2$ were
available, as is the case of CS, we scaled those calculated for He
by multiplying them by the squared ratio of the reduced masses of
the H$_2$ and He colliding systems. When rate coefficients were
available for both ortho and para H$_2$, an ortho/para ratio of 3
was assumed. Wherever we needed rate coefficients involving
rotational levels higher than those included in the quantum
calculations, we computed them by using the Infinite Order Sudden
(IOS) approximation (\cite{gol1977} 1977). No extrapolation in
temperature has however been made outside the temperature validity
range of the quantum calculations. Finally, the collision
excitation rates have been systematically computed from the
de-excitation rates applying detailed balance. More details are
given below for individual molecules.

-- CS: we have considered the first 50 rotational levels within
the vibrational states $v$ = 0, 1, 2, 3. The level energies and
transition frequencies were calculated from the Dunham
coefficients provided by \cite{mul2005} (2005), the line strengths
for rotational transitions were computed from the dipole moments
for each vibrational state $\mu_{v=0}$ = 1.985 D, $\mu_{v=1}$ =
1.936 D (\cite{win1968} 1968), $\mu_{v=2}$ = 1.914 D
({\footnotesize CDMS}\footnote{See
\texttt{http://www.astro.uni-koeln.de/cdms/} \label{foot-cdms}}),
and $\mu_{v=3}$ = 1.855 D (\cite{loptip1987} 1987), while for
ro-vibrational transitions the Einstein coefficients were obtained
from the values calculated by \cite{cha1995} (1995) for the P(1)
transition of each vibrational band (see e.g. \cite{tip1981}
1981). Rate coefficients for de-excitation through inelastic
collisions with H$_2$ and He were taken from \cite{liq2006}
(2006), who calculated coefficients for collisions with He
including the first 31 rotational levels and up to temperatures of
300 K, and from \cite{liq2007} (2007), who included the first 38
rotational levels within the first 3 vibrational states and
computed collision coefficients for rotational and ro-vibrational
transitions at temperatures between 300 and 1500 K.

-- SiO: the first 50 rotational levels of the $v=0$ and $v=1$
vibrational states have been included. The Dunham coefficients
given by \cite{san2003} (2003), the dipole moments $\mu_{v=0}$ =
3.0982 D and $\mu_{v=1}$ = 3.1178 D measured by \cite{ray1970}
(1970), and the Einstein coefficient for the $v=1-0$ P(1)
transition calculated by \cite{dri1997} (1997) were used to
compute the line frequencies and strengths. The collision rate
coefficients calculated by \cite{day2006} (2006) for the first 20
rotational levels and for temperatures up to 300 K were adopted.
For temperatures higher than 300 K and for ro-vibrational
transitions we adopted the collision rate coefficients used for
CS.

-- SiS: we have included the first 70 rotational levels within the
$v$ = 0, 1, 2, 3, 4, 5 vibrational states. The Dunham coefficients
and the dipole moment of 1.735 D given by \cite{mul2007} (2007),
together with the dipole moments for ro-vibrational transitions
given by \cite{lop1987} (1987) have been used to compute level
energies and Einstein coefficients. The rate coefficients of
rotational de-excitation through collisions with He and ortho/para
H$_2$ have been taken from the calculations of \cite{vin2007}
(2007) and \cite{klo2008} (2008), which extend up to temperatures
of 200--300 K. For higher temperatures and for ro-vibrational
transitions the collision rate coefficients calculated by
\cite{tob2008} (2008) were used.

-- Metal halides: the spectroscopic properties of NaCl, KCl, AlCl,
and AlF are relatively well known (\cite{car2002} 2002, 2004,
\cite{hed1993} 1993, \cite{hed1992} 1992). The permanent electric
dipole moment is very large for NaCl (9.00117 D; \cite{del1970}
1970) and KCl (10.269 D; \cite{van1967} 1967), and noticeably
smaller for AlCl and AlF, 1--2 D and 1.53 D, respectively
(\cite{lid1965} 1965). For AlCl we adopted a value of 1.5 D, in
agreement with the estimation of \cite{lid1965} (1965) and the
value previously used by \cite{cer1987} (1987). We have not
included vibrationally excited states since no lines involving
such states are observed in IRC +10216, and since vibrational line
strengths for those molecules are poorly known. Recently,
\cite{got2011} (2011,2012) have calculated the rate coefficients
for AlF de-excitation through collisions with He and para H$_2$ up
to temperatures of 300 and 70 K, respectively. We have also used
those rate coefficients for AlCl, which has a similar dipole
moment than AlF. For NaCl and KCl, we have adopted the values used
for SiS, which extend up to higher temperatures. The lack of
accurate collision rate coefficients at high temperatures for
these metal halides introduce significant uncertainties in the
derived abundances.

-- NaCN: we consider rotational levels within the ground
vibrational state up to $J=29$. The rotational constants used to
compute the level energies were derived from a fit to the line
frequencies measured in the spectrum of IRC +10216 and in the
laboratory by \cite{hal2011} (2011)$\footnote{We note that the
rotational constants in Table~2 of \cite{hal2011} (2011) are
innacurate, as they result in line frequencies which differ by
several MHz from the measured ones.}$. The dipole moment of NaCN
is calculated to be 8.85 D ({\footnotesize
CDMS}$^{\ref{foot-cdms}}$). Due to the lack of calculated
collision data for NaCN we use the following approximate formula
to compute the de-excitation rate coefficients $\gamma$ due to
collisions with H$_2$:
\begin{equation}
\log \gamma = -10 - 0.2 \times (J' - J'') - 0.4 \times (K_a' -
K_a'') \label{eq-approximate-collision-coeffs}
\end{equation}
where $'$ and $''$ stand for upper and lower level, respectively,
and $\gamma$ has units of cm$^3$ s$^{-1}$.
Eq.~(\ref{eq-approximate-collision-coeffs}) gives a first
approximation of collision rate coefficients for asymmetric
rotors, resulting in values within one order of magnitude of those
calculated for asymmetric rotors such as SO$_2$ (\cite{gre1995}
1995) and HCO$_2^+$ (\cite{ham2007} 2007). We, nevertheless, note
that the use of this approximate formula may introduce significant
errors in the excitation and abundance of NaCN.

The lack of collision rate coefficients for AlCl, NaCl, KCl, and
NaCN may introduce important uncertainties in the derived
abundances. We may in fact question whether the choice of the
collision rate coefficients for these molecules is more adequate
than the simple assumption of local thermodynamic equilibrium
(LTE). In the case of AlCl, the choice of the collision
coefficients calculated for AlF does not seem a bad approach based
on the similar dipole moment and mass of both molecules. In any
case, the observed lines of AlCl (see discussion in
Sec.~\ref{subsec-radiative-transfer-results}) arise from regions
where the level populations are very close to thermalization, and
therefore both the use of collision coefficients and the LTE
approach yield similar results. This is not the case for NaCl,
KCl, and NaCN (see also discussion in
Sec.~\ref{subsec-radiative-transfer-results}), for which the
assumption of LTE results in a very bad agreement with the
observed lines (the calculated low--$J$ lines are much too weak
and the high--$J$ lines are too strong). We are therefore
confident that our choice of the collision rate coefficients for
NaCl, KCl, and NaCN yields more reliable abundances than the mere
assumption of LTE.

\subsection{Modeling strategy}
\label{subsec-modeling-strategy}

Once the physical parameters of the envelope, described in
Sec.~\ref{sec-physical-model}, and the spectroscopic data and
collision rate coefficients, detailed in
Sec.~\ref{subsec-molecular-data}, have been established, the only
remaining parameter needed to compute the emergent line profiles
is the abundance radial profile for each of the studied molecules.

As a first trial, we start with a constant abundance from the
stellar photosphere up to the outer layers where molecules are
photodissociated by interstellar ultraviolet photons. The
abundance fall off due to photodissociation is calculated with a
simple chemical model where the photodissociation rates are
parameterized as a function of the visual extinction $A_V$.
Photodissociation rates for CS and SiO (this latter adopted also
for SiS) are taken from \cite{van2006} (2006), for NaCl (adopted
also for KCl and AlCl) it is taken from \cite{van1998} (1998),
while for AlF and NaCN we have adopted the educated guess
10$^{-9}$ s$^{-1}$ $\exp(-1.7 \times A_V)$. The visual extinction
at each radius $r$ is derived by the N$_{\rm H}$/$A_V$ ratio, 2.7
$\times$ 10$^{21}$ cm$^{-2}$, derived from our adopted dust
parameters and gas-to-dust mass ratio, where N$_{\rm H}$ is the
column density of total hydrogen nuclei from the radius $r$ up to
the end of the envelope. The intensity adopted for the local
interstellar radiation field is half that of the standard field of
\cite{dra1978} (1978); see a discussion on this point in
\cite{agu2006} (2006). Once the abundance fall-off due to
photodissociation is determined, line profiles are calculated and
compared to the observations. Then, for each molecule, the initial
abundance is progressively varied until the calculated line
profiles match the observed ones.

A constant abundance profile does not always yield a satisfactory
fit to the observed line profiles. More specifically, for some
molecules it is found that the calculated intensity is
overestimated for low--$J$ lines and underestimated for high--$J$
lines, implying a larger fractional abundance in the warm inner
layers than in the cold outer regions. Such abundance gradient,
likely caused by condensation on grains, has been calculated
through a simple chemical model which considers that from the
condensation radius $R_c$ gas phase molecules may me adsorbed onto
grain surfaces, with a rate proportional to a sticking
coefficient, the thermal velocity of the molecules, the
geometrical section of the dust grains, and the number of dust
grains per unit volume. All such parameters are fixed by the
physical model described in Sec.~\ref{sec-physical-model}, except
for the sticking coefficient which is varied for each molecule
until the calculated low--$J$ and high--$J$ line profiles agree
with the observed ones. The radial abundance profiles obtained in
this way show a pronounced gradient after $R_c$ and become nearly
flat at a radius of about 2 $\times$ 10$^{15}$ cm. This approach
does not pretend to precisely simulate the condensation process,
but at least yields a smooth abundance transition between the
inner and the outer layers of the envelope and gives satisfactory
results.

Still, with the approach described above, the calculated high--$J$
lines of some metal-bearing molecules show extra emission at the
line center, compared with the observed line profiles. This
implies that the innermost regions, where the expansion velocity
is small, contribute too much in the model to the overall line
emission. In such cases we have allowed for a step--like abundance
decrease in the innermost regions, from the photosphere up to
about 3 R$_*$, where thermochemical equilibrium prevails (see
\cite{agu2006} 2006).

The best fit model, and so the final abundance profile, for each
molecule is chosen on the basis of the best overall agreement
between the calculated and the observed line profiles. The
obtained abundance profiles are further discussed for individual
molecules in Sec.~\ref{subsec-molecular-abundances}.

\subsection{Results of the radiative transfer modeling}
\label{subsec-radiative-transfer-results}

\begin{figure}
\includegraphics[angle=0,scale=.47]{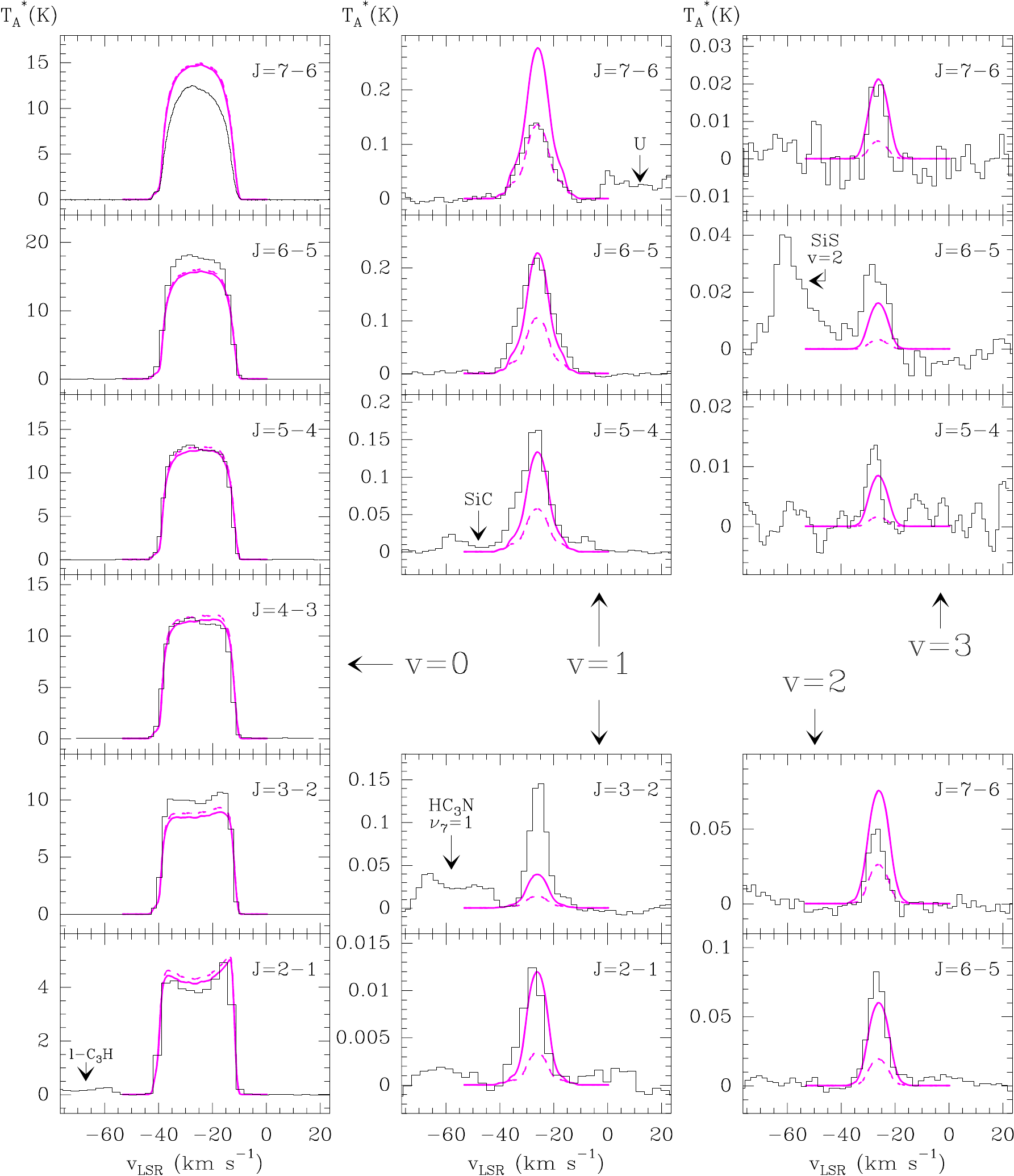}
\caption{Rotational lines of CS in IRC +10216 as observed with the
IRAM 30-m telescope (black histograms) and as calculated with the
radiative transfer model (magenta lines). Continuous lines refer
to the best model, using the abundance profile shown in
Fig.~\ref{fig-abun-cs}, while dashed lines correspond to a model
with a constant abundance of 7 $\times$ 10$^{-7}$ from 1 R$_*$ up
to the photodissociation region. Note that this latter model
underestimates the intensity of vibrationally excited lines.}
\label{fig-cs}
\end{figure}

\begin{figure}
\begin{center}
\includegraphics[angle=0,scale=.47]{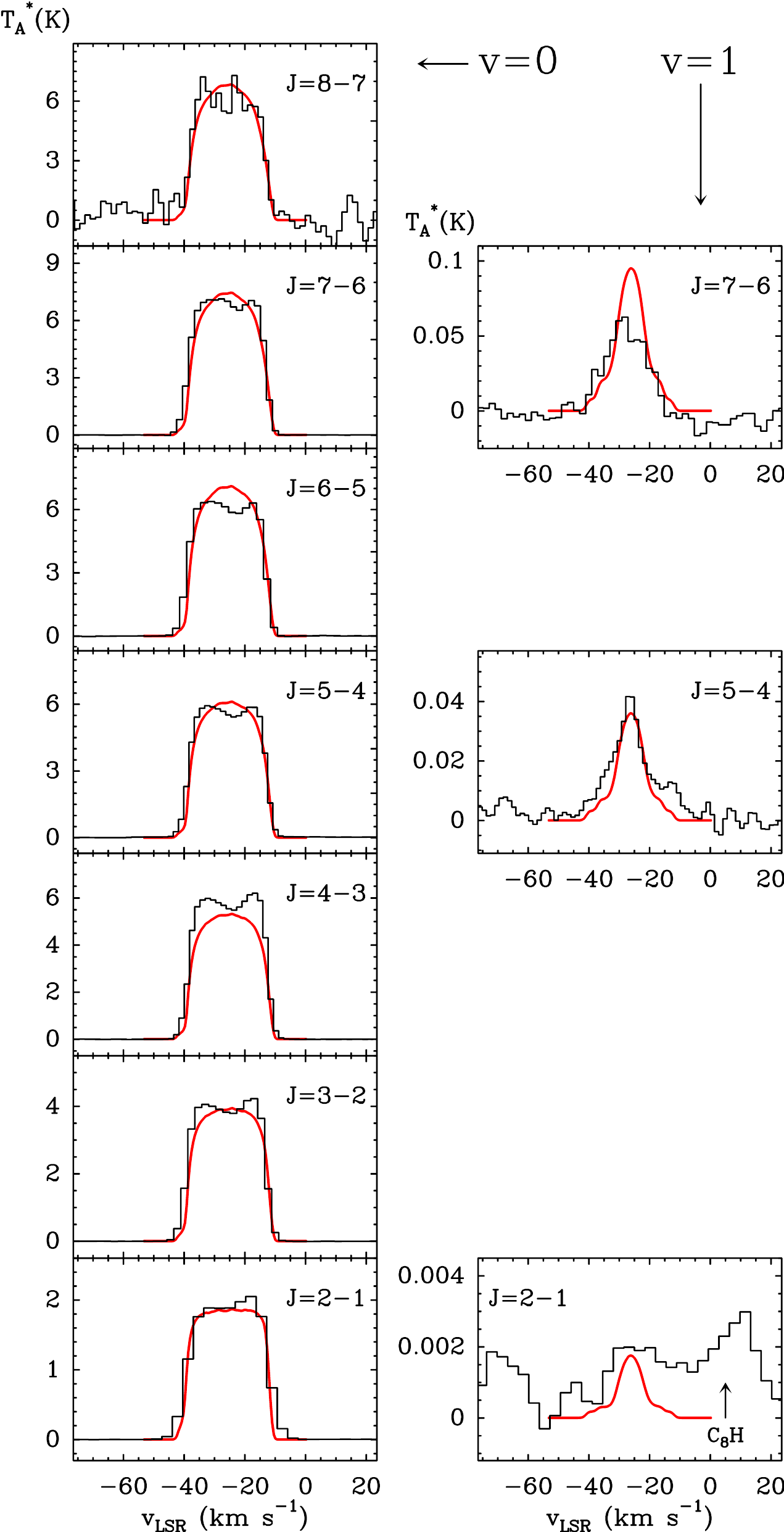}
\caption{Rotational lines of SiO in IRC +10216 as observed with
the IRAM 30-m telescope (black histograms) and as calculated with
the radiative transfer model (red lines).} \label{fig-sio}
\end{center}
\end{figure}

\begin{figure*}
\begin{center}
\includegraphics[angle=0,scale=.88]{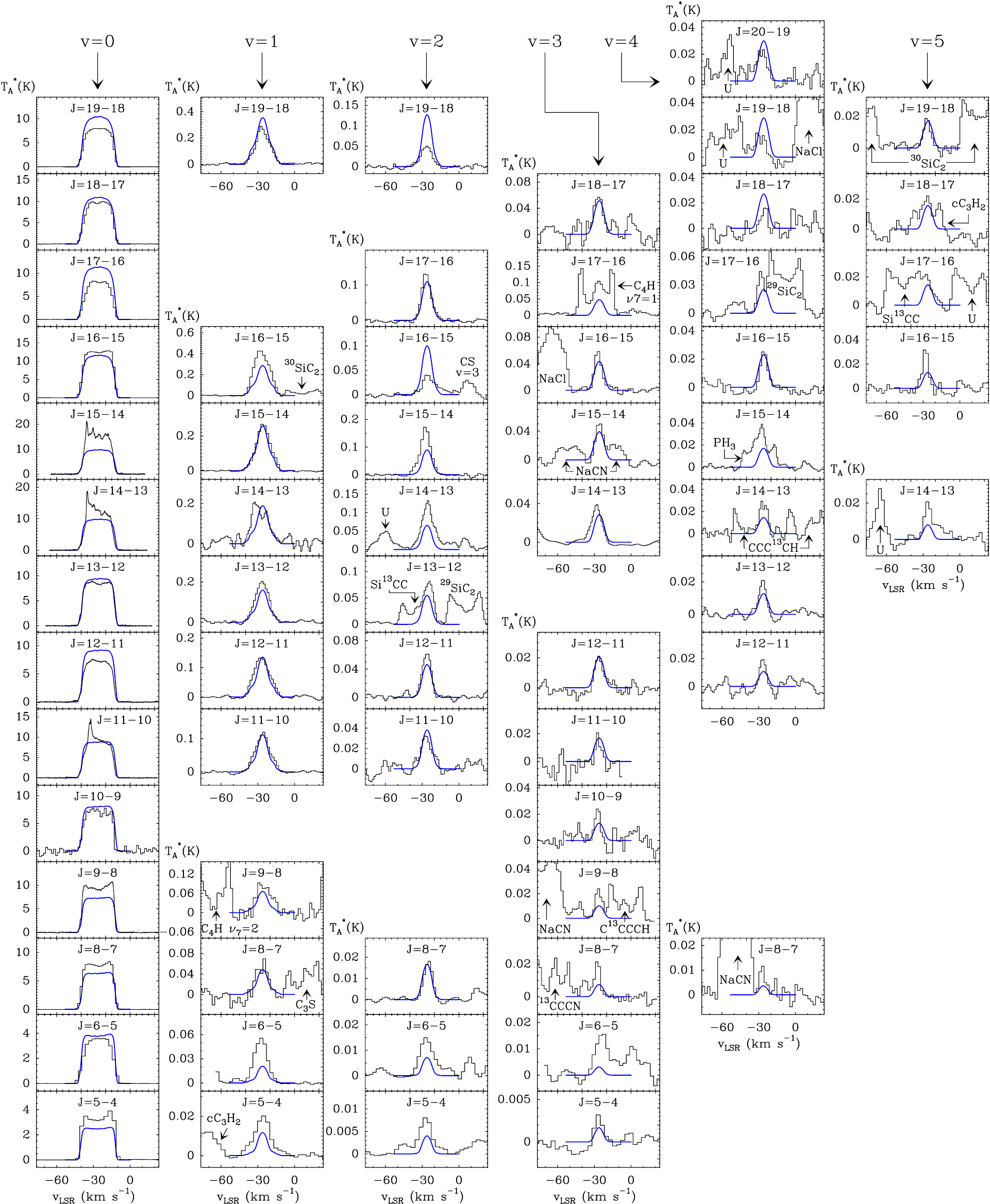}
\caption{Rotational lines of SiS in IRC +10216 as observed with
the IRAM 30-m telescope (black histograms) and as calculated with
the radiative transfer model (blue lines).} \label{fig-sis}
\end{center}
\end{figure*}

The observed rotational lines of CS, SiO, and SiS are plotted in
Figs.~\ref{fig-cs}, \ref{fig-sio}, and \ref{fig-sis},
respectively, together with the line profiles calculated through
the radiative transfer model. In Fig.~\ref{fig-abun-cs} we plot
the derived abundance profiles. For these three molecules the
observed lines cover a wide range of upper level energies, up to
5500 K for CS (level $v=3, J=7$), 1830 K for SiO (level $v=1,
J=7$), and 5440 K for SiS (level $v=5, J=19$), so that they sample
different regions of the envelope. Lines from the ground
vibrational state trace the cool mid and outer layers while
vibrationally excited lines trace the warm material located in the
innermost regions ($r \lesssim 5$ R$_*$). The reader may note that
line widths of vibrationally excited states are significantly
narrower than those of the ground vibrational state, indicating
that the emission in these lines comes from the innermost regions
where the gas has not been yet accelerated up to the terminal
expansion velocity of 14.5 km s$^{-1}$.

As shown in Fig.~\ref{fig-cs}, there is a good overall agreement
between calculated and observed line profiles for CS, except for
the $v=1$, $J=3-2$ line whose observed intensity is substantially
larger than calculated. This line is probably affected by a maser
amplification, as indicated by the high spectral resolution
observations made by \cite{hig2000} (2000). These authors point
towards a mechanism based on infrared pumping to the $v=1$ state,
something that is included in our radiative transfer treatment,
but does not produce a maser amplification in this line. A more
likely mechanism for this relatively weak maser could be the
overlap of ro-vibrational CS lines with those of abundant
molecules such as SiO, a mechanism that explain the maser emission
observed in oxygen-rich AGB stars in several rotational
transitions of $^{29}$SiO and $^{30}$SiO, and of SiO in the $v=3$
and $v=4$ vibrational states (\cite{gon1997} 1997), and in some
SiS rotational lines within the ground vibrational state in IRC
+10216 (\cite{fon2006} 2006; see below).

The excitation of CS throughout the envelope is dominated by
collision processes although the absorption of infrared photons
emitted by the central star and by the warm dust also plays an
important role. Radiative pumping at 8 $\mu$m populates the
vibrationally excited states, enhancing the emission in lines such
as the $v=3$ $J=7-6$ (whose levels lie 5500 K above the ground
state and which otherwise would not be detectable), but also,
through downward cascades, affects the population of the $v=0$
rotational levels. Therefore, a proper treatment of infrared
pumping warrants a correct estimation of the abundance in the
inner layers (through the vibrationally excited lines) and allows
also to accurately determine the abundance in the outer envelope,
which would be overestimated by about 30 \% if IR pumping were
neglected. According to our analysis, the abundance of CS relative
to H$_2$ is 4 $\times$ 10$^{-6}$ in the inner layers and drops to
7 $\times$ 10$^{-7}$ in the outer layers (see
Fig.~\ref{fig-abun-cs}). The estimated uncertainty of the
abundance is 50 \% in the outer envelope, and a factor of 2 for
the regions inner to R$_c$, where densities are more uncertain.
Fig.~\ref{fig-cs} shows the improvement in the fit to the line
profiles brought by the two abundance components model.

In the case of SiO, we have observed the $J=2-1$ through $J=7-6$
lines of the ground vibrational state as well as 3 lines of the
$v=1$ state. As shown in Fig.~\ref{fig-sio}, there is a good
agreement between calculated and observed line profiles. A slight
difference is however found in the $v=0$ line shapes (calculated
are parabolic while observed have a slight double-peak character),
which may be indicative of a more extended SiO envelope. A better
agreement is found, rather than by extending SiO to larger radii,
by enhancing its excitation in the outer layers. A possible source
for the required additional excitation could be provided by shells
with enhanced density, known to be present in the outer layers of
IRC +10216 (see \cite{mau1999} 1999; \cite{cor2009} 2009;
\cite{deb2012} 2012), or by an infrared pumping larger than given
by our model. We have nevertheless not attempted to improve the
agreement for SiO, which overall is satisfactory enough, by
modifying the physical model, since this would affect all the
other molecules. The excitation of SiO levels in the model is very
similar to that found for CS, i.e., is controlled by inelastic
collisions with H$_2$ while radiative pumping at 8 $\mu$m affects
the intensities of the $v=0$ rotational lines by about 30 \% and
completely dominates the excitation of the vibrationally excited
states. The derived SiO abundance, 1.8 $\times$ 10$^{-7}$ relative
to H$_2$, remains constant from the innermost layers up to the
photodissociation region (see Fig.~\ref{fig-abun-cs}). The
uncertainty in the abundance is estimated to be 50 \% in the outer
layers, rising to a factor of 2 in the inner regions, which are
sampled by just a few vibrationally excited lines and where
densities are more uncertain.

Silicon monosulphide (SiS) has been observed in a large number of
rotational lines covering the $v=0$ through $v=5$ vibrational
states (see Fig.~\ref{fig-sis}). The analysis of this molecule is
complicated by the fact that some rotational lines within the
ground vibrational state (concretely the $J=11-10$, $J=14-13$, and
$J=15-14$) show a profile with enhanced emission due to maser
amplification at certain velocities, which are probably caused by
the overlap at 13.5 $\mu$m of ro-vibrational transitions of SiS
with those of C$_2$H$_2$ and HCN (\cite{fon2006} 2006). The
agreement between calculated and observed line profiles is
somewhat poorer than in the cases of CS and SiO. On the one hand
our radiative transfer calculations do not include infrared
overlaps, something that clearly affects the three maser lines
previously mentioned but can also have an influence on the
excitation of other vibrationally excited lines (see
\cite{fon2006} 2006), and on the other hand the observed intensity
of some high frequency lines may have non negligible errors due to
calibration (estimated to be around 30 \% in the $\lambda$0.9 mm
band), to pointing inaccuracies (which are typically $<$2$''$,
although the main beam of the IRAM 30-m telescope is as small as
7$''$ at 360 GHz and vibrationally excited lines have an emission
size of less than 1$''$), or even due to time variability
(difficult to evaluate due to the lack of observations at
different times in the $\lambda$0.9 mm band).

Infrared pumping affects little the level populations within the
ground vibrational state of SiS, but controls, at practically the
same extent than collisions, the excitation of the vibrationally
excited lines. As in the case of CS, we need to increase the
abundance of SiS in the inner layers with respect to that in the
outer envelope. We derive for SiS an abundance relative to H$_2$
of 3 $\times$ 10$^{-6}$ in the inner layers, decreasing down to a
value of 1.3 $\times$ 10$^{-6}$ in the mid and outer layers (see
Fig.~\ref{fig-abun-cs}). The estimated uncertainty in the
abundance of SiS is a factor of 2 in the inner layers and 60 \%
for the outer envelope.\\

\begin{figure}
\begin{center}
\includegraphics[angle=0,scale=.55]{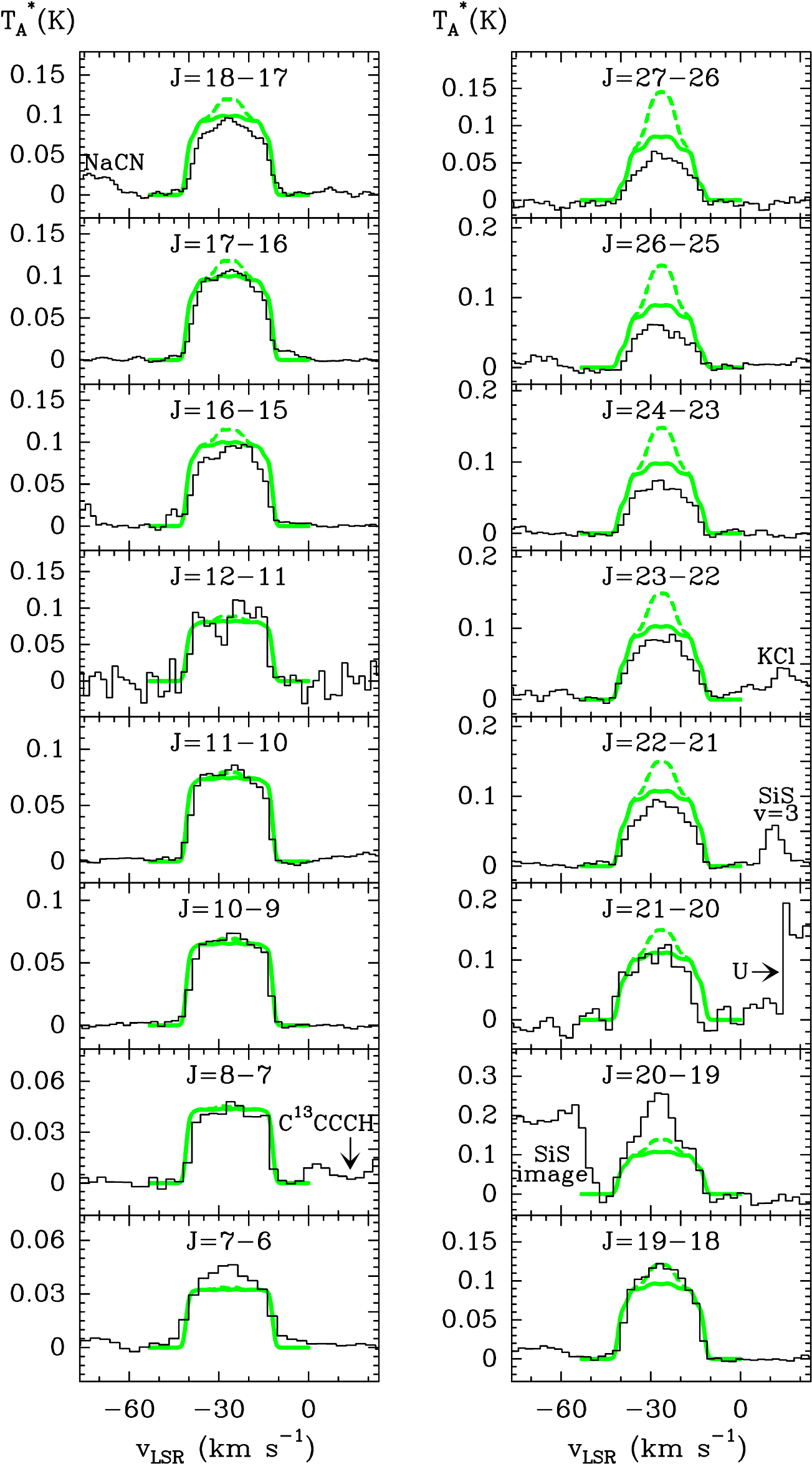}
\caption{Rotational lines of Na$^{35}$Cl in IRC +10216 as observed
with the IRAM 30-m telescope (black histograms) and as calculated
with the radiative transfer model (green lines). Continuous lines
refer to the best model, using the abundance profile shown in
Fig.~\ref{fig-abun-nacl}, while dashed lines result from a model
without the abundance decrease in the 1--3 R$_*$ region. Note that
dashed lines overestimate the intensity at the center of high--$J$
lines.} \label{fig-nacl}
\end{center}
\end{figure}

\begin{figure}
\begin{center}
\includegraphics[angle=0,scale=.52]{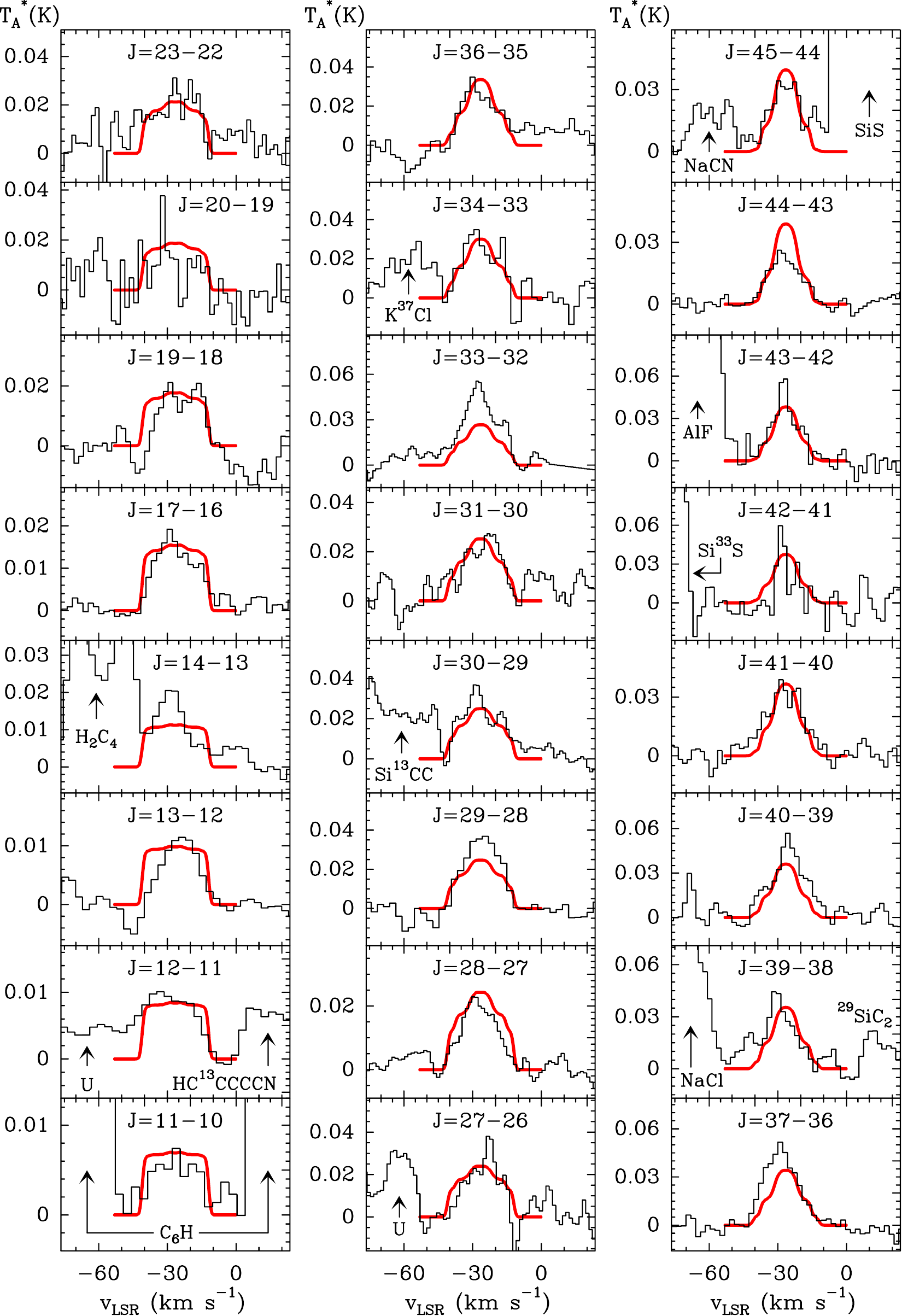}
\caption{Rotational lines of K$^{35}$Cl in IRC +10216 as observed
with the IRAM 30-m telescope (black histograms) and as calculated
with the radiative transfer model (red lines).} \label{fig-kcl}
\end{center}
\end{figure}

The metal halides NaCl, KCl, AlCl, and AlF are all observed
through rotational transitions within the ground vibrational
state. No vibrationally excited lines are detected, probably
because of the low abundance of these species. The observed and
calculated line profiles of Na$^{35}$Cl and K$^{35}$Cl are shown
in Fig.~\ref{fig-nacl} and Fig.~\ref{fig-kcl}, respectively. These
two species have similar dipole moments, so that the larger
intensity of the NaCl lines largely reflects its larger abundance
with respect to KCl. The abundance derived for Na$^{35}$Cl in the
inner layers is 1.3 $\times$ 10$^{-9}$ relative to H$_2$ (with an
uncertainty of a factor of 2), although we find necessary to
decrease the abundance in the thermochemical equilibrium region
(1--3 R$_*$) down to a value of 10$^{-11}$ (see
Fig.~\ref{fig-abun-nacl}) to avoid a too strong emission for the
high--$J$ lines near the central velocity (see
Fig.~\ref{fig-nacl}). In the case of K$^{35}$Cl the derived
abundance profile is somewhat more elaborated (see
Fig.~\ref{fig-abun-nacl}). KCl abundance is found to have a slight
gradient going from 5 $\times$ 10$^{-10}$ at 5 R$_*$ down to 3.7
$\times$ 10$^{-10}$ in the outer layers, with an estimated error
of a factor of 2. As with NaCl, the abundance is decreased in the
1--3 R$_*$ region, although in this case by just a factor of 2.5.

\begin{figure}
\begin{center}
\includegraphics[angle=0,scale=.55]{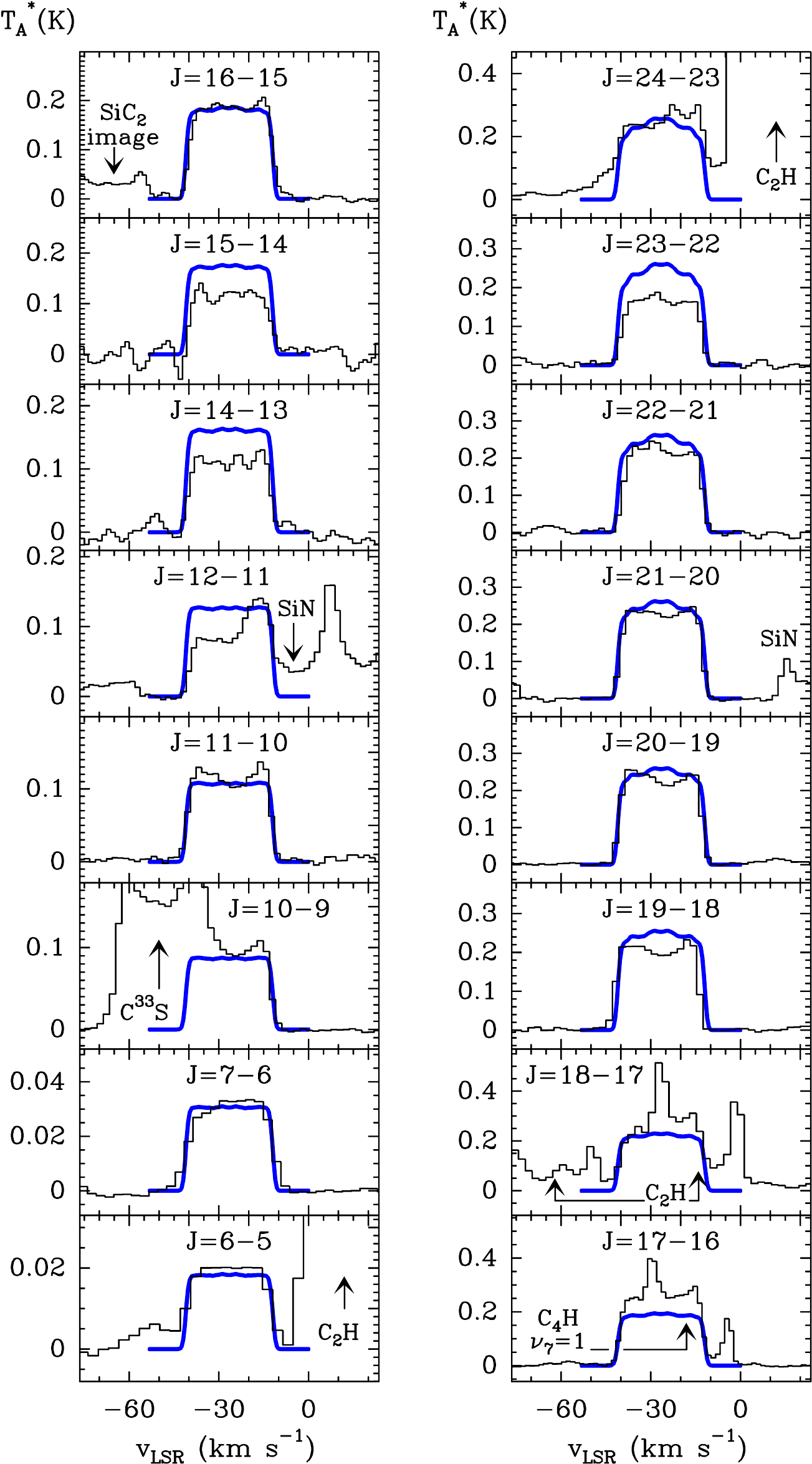}
\caption{Rotational lines of Al$^{35}$Cl in IRC +10216 as observed
with the IRAM 30-m telescope (black histograms) and as calculated
with the radiative transfer model (blue lines).} \label{fig-alcl}
\end{center}
\end{figure}

\begin{figure}
\begin{center}
\includegraphics[angle=0,scale=.43]{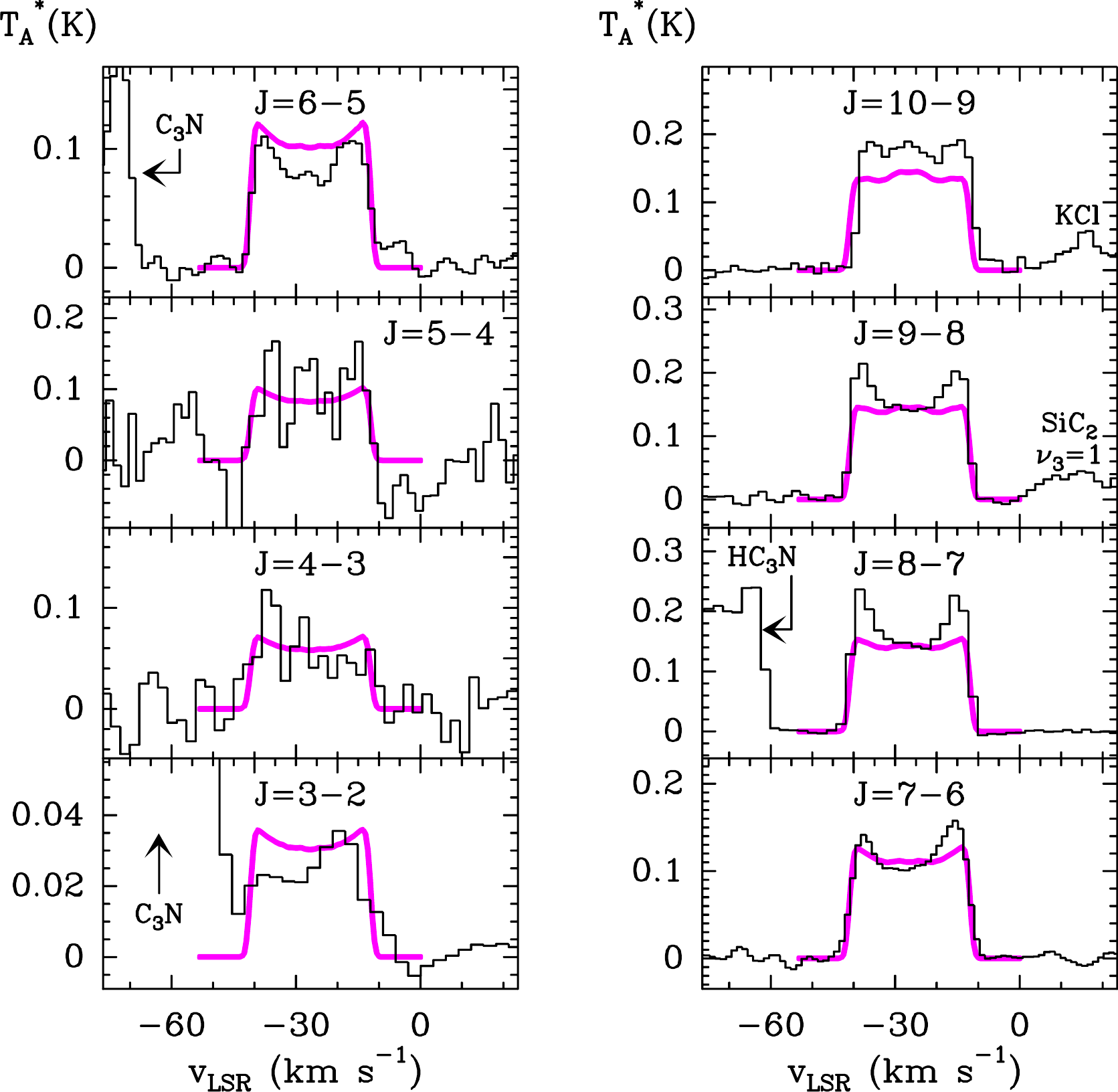}
\caption{Rotational lines of AlF in IRC +10216 as observed with
the IRAM 30-m telescope (black histograms) and as calculated with
the radiative transfer model (magenta lines).} \label{fig-alf}
\end{center}
\end{figure}

The observed and calculated line profiles of Al$^{35}$Cl and AlF
are shown in Fig.~\ref{fig-alcl} and Fig.~\ref{fig-alf},
respectively. AlCl and AlF have a much smaller dipole moment than
NaCl, but show similar line intensities, which suggest they are
fairly more abundant. Radiative transfer calculations confirm this
point and yield abundances relative to H$_2$ of 5 $\times$
10$^{-8}$ for Al$^{35}$Cl and 10$^{-8}$ for AlF (with an
uncertainty of 50 \% increasing up to a factor of 2 in the regions
inner to $R_c$). In the thermochemical equilibrium region (1--3
R$_*$) the abundance of Al$^{35}$Cl is decreased, as for
Na$^{35}$Cl and K$^{35}$Cl, by a factor of 5 (see
Fig.~\ref{fig-abun-nacl}). It is worth to note that the observed
lines of AlF are slightly more U-shaped than those of AlCl. Such
behavior can be accounted for to some extent by assuming that AlF
is more extended than AlCl due to a lower photodissociation rate,
as assumed in our model (see Sec.~\ref{subsec-modeling-strategy}).
Still the observed line profiles of AlF have a slightly more
pronounced U-shape than calculated (see Fig.~\ref{fig-alf}). Here,
as occurs with SiO (see above), a better agreement cannot be
obtained by extending AlF to larger radii, but by enhancing the
excitation of AlF in the outer layers. In fact, assuming LTE
excitation for AlF brings the calculated line profiles in close
agreement with the observed ones. This is somewhat paradoxical
since AlF is the only metal-bearing molecule for which collision
rate coefficients have been calculated. This slight discrepancy
is, nevertheless, unlikely to have a significant impact on the AlF
derived abundance.

The excitation of the metal halides in IRC +10216 is likely to be
dominated by inelastic collisions. The role of infrared pumping to
vibrationally excited levels, not included in our radiative
transfer calculations, is difficult to evaluate mostly due to the
lack of ro-vibrational line strengths. To get a rough idea of the
importance of infrared pumping we have run radiative transfer
models including the first vibrationally excited state. As dipole
moment for the $v$=1$\rightarrow$0 band we have adopted that of
SiS (0.13 D), scaled by the ratio of the halide to SiS permanent
electric dipole moments. These calculations indicate that infrared
pumping does not significantly affect the intensities of the
observed lines, except for the high$-J$ lines of NaCl which are
slightly more intense with respect to the case when infrared
pumping is not included. We may conclude that it is unlikely that
infrared pumping plays an important role in the excitation of the
$v=0$ rotational levels of these metal halides, which is very
likely dominated by collision processes.

\begin{figure}[b]
\begin{center}
\includegraphics[angle=0,scale=.37]{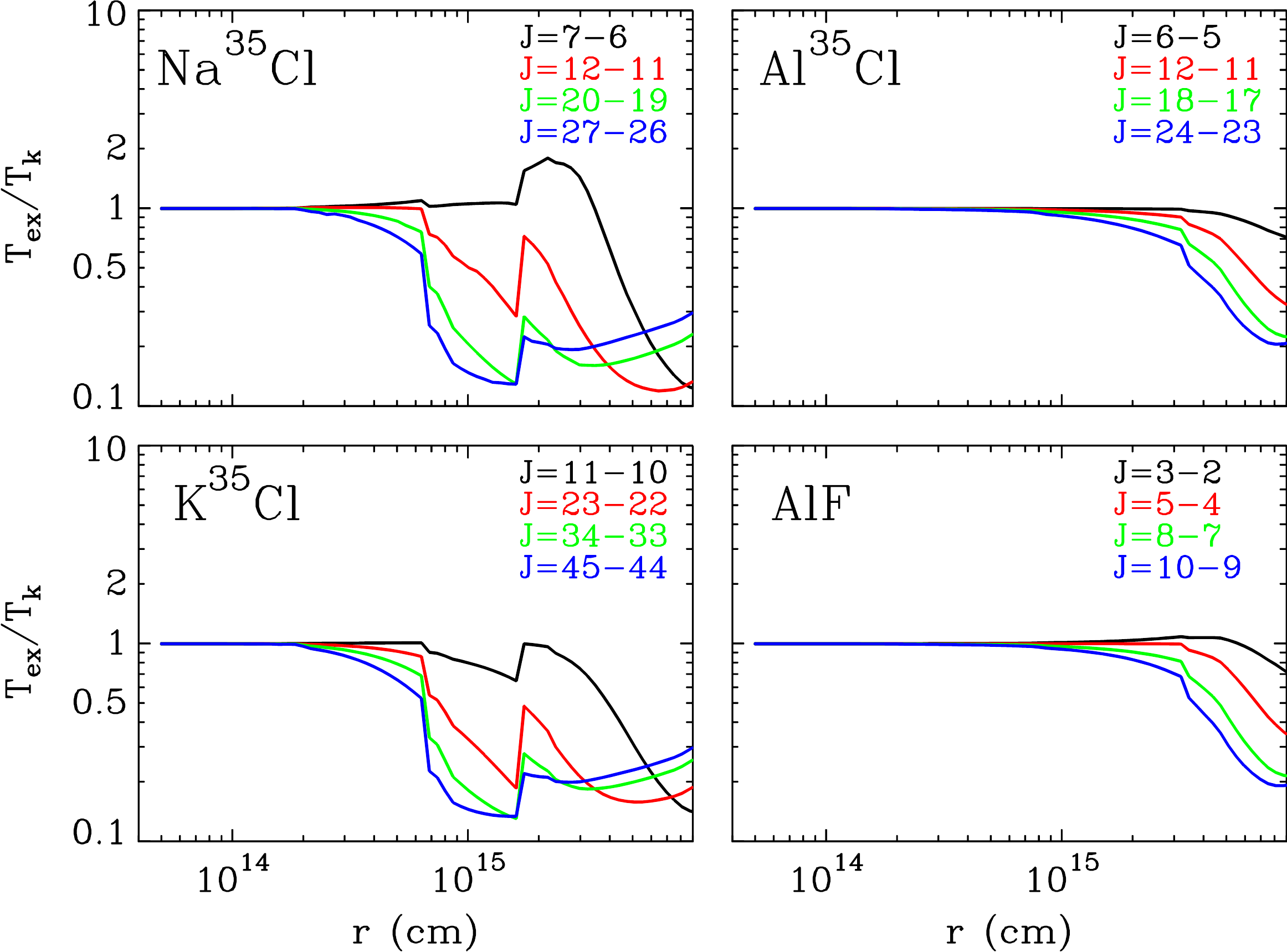}
\caption{Calculated ratio of excitation temperature to kinetic
temperature ($T_{\rm ex}$ / $T_k$) as a function of radius for
selected rotational transitions of the metal halides Na$^{35}$Cl,
K$^{35}$Cl, Al$^{35}$Cl, and AlF in IRC +10216.}
\label{fig-tex-metalhalides}
\end{center}
\end{figure}

Even in the absence of IR pumping, the ground vibrational state
populations of metal halides depart from LTE, according to our
statistical equilibrium calculations.
Fig.~\ref{fig-tex-metalhalides} shows, for selected rotational
transitions, the calculated ratio of the excitation to kinetic
temperature as a function of radius. We see that for the 4 metal
halides this ratio is equal to 1 in the inner layers, i.e. the
rotational levels are thermalized, whereas at larger radii the
decrease in density makes rotational levels to be subthermally
populated yielding excitation temperatures lower than the kinetic
temperature. NaCl and KCl, with a large dipole moment, start to
deviate from thermalization at radii much shorter than AlCl and
AlF, which due to their moderately low dipole moments keep close
to thermalization throughout a large part of the envelope. This
suggests that the choice of the collision rate coefficients should
not be critical for AlCl and AlF.

\begin{figure*}
\begin{center}
\includegraphics[angle=0,scale=0.95]{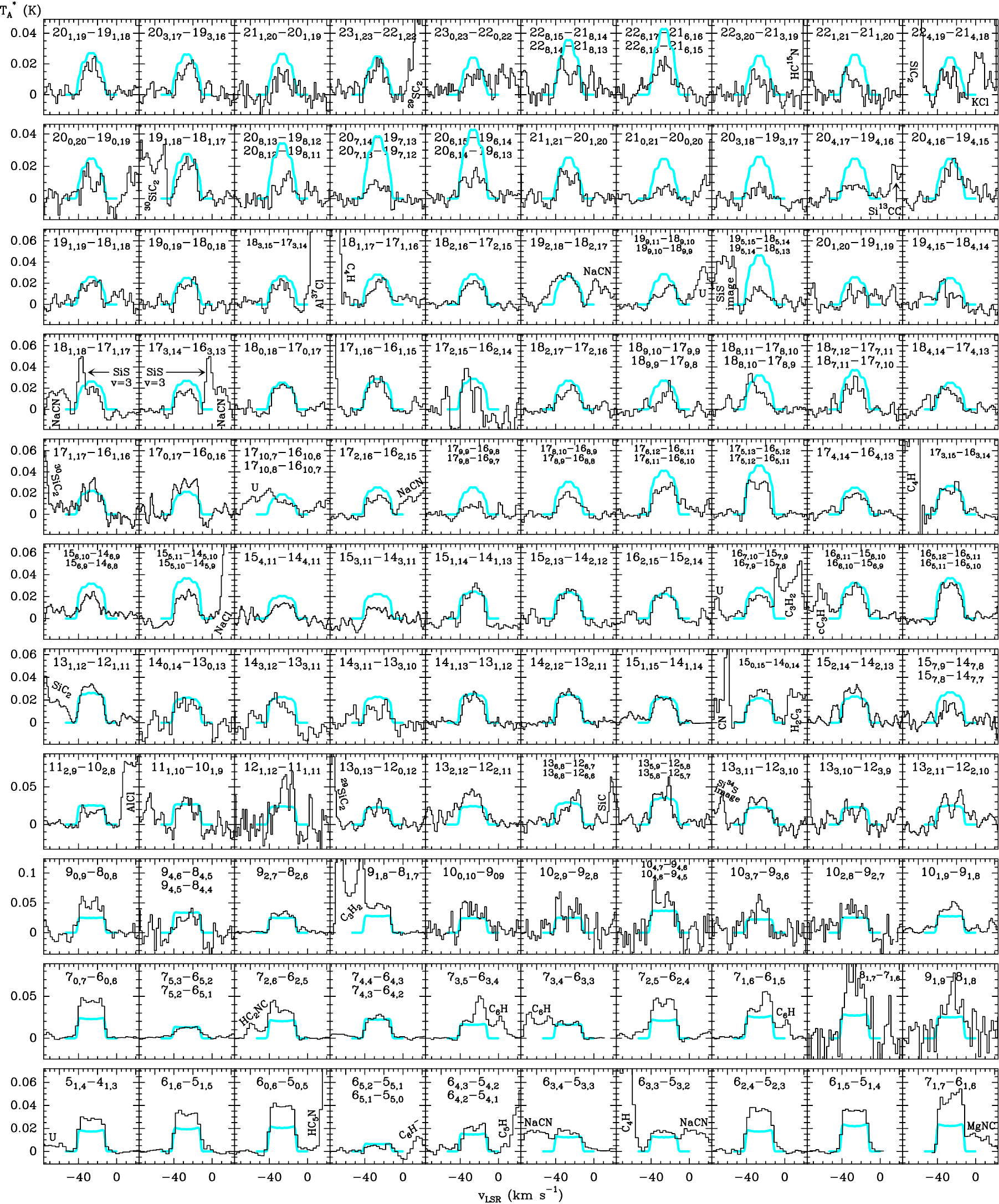}
\caption{Rotational lines of NaCN in IRC +10216 as observed with
the IRAM 30-m telescope (black histograms) and as calculated with
the radiative transfer model (blue lines). Lines are ordered by
increasing frequency from bottom to top (and from left to right).}
\label{fig-nacn}
\end{center}
\end{figure*}

Finally, as concerns the cyanide NaCN, a fairly large number of
rotational lines within the ground vibrational state has been
observed (see Fig.~\ref{fig-nacn}). Although the agreement between
calculated and observed line profiles is overall satisfactory,
taking into account the large uncertainty of the collision rate
coefficients, the calculated high$-J$ high$-K_a$ lines are too
strong and the low$-J$ and low$-K_a$ lines too weak. We note that
the discrepancy is larger if we replace statistical equilibrium
calculations by LTE. The adopted photodissociation rate of NaCN is
also very uncertain (see Sec.~\ref{subsec-modeling-strategy}). Our
model points toward a lower photodissociation rate than adopted,
which would extend the NaCN envelope increasing the intensity of
low$-J$ lines, although this conclusion is hampered by the
uncertain collision rates adopted, which are likely to cause most
of the observed discrepancies. The role of infrared pumping to
vibrationally excited states is difficult to evaluate due to the
lack of spectroscopic and line strengths data for such states. The
derived abundance for NaCN is 3 $\times$ 10$^{-9}$ relative to
H$_2$, with an estimated uncertainty of a factor of 3. Decreasing
the NaCN abundance in the 1--3 R$_*$ region, down to a very low
value (10$^{-12}$), improves somewhat the fit to the observed
high$-J$ high$-K_a$ lines.

\section{Discussion} \label{sec-discussion}

\subsection{Molecular abundances}
\label{subsec-molecular-abundances}

\begin{figure}
\begin{center}
\includegraphics[angle=0,scale=.43]{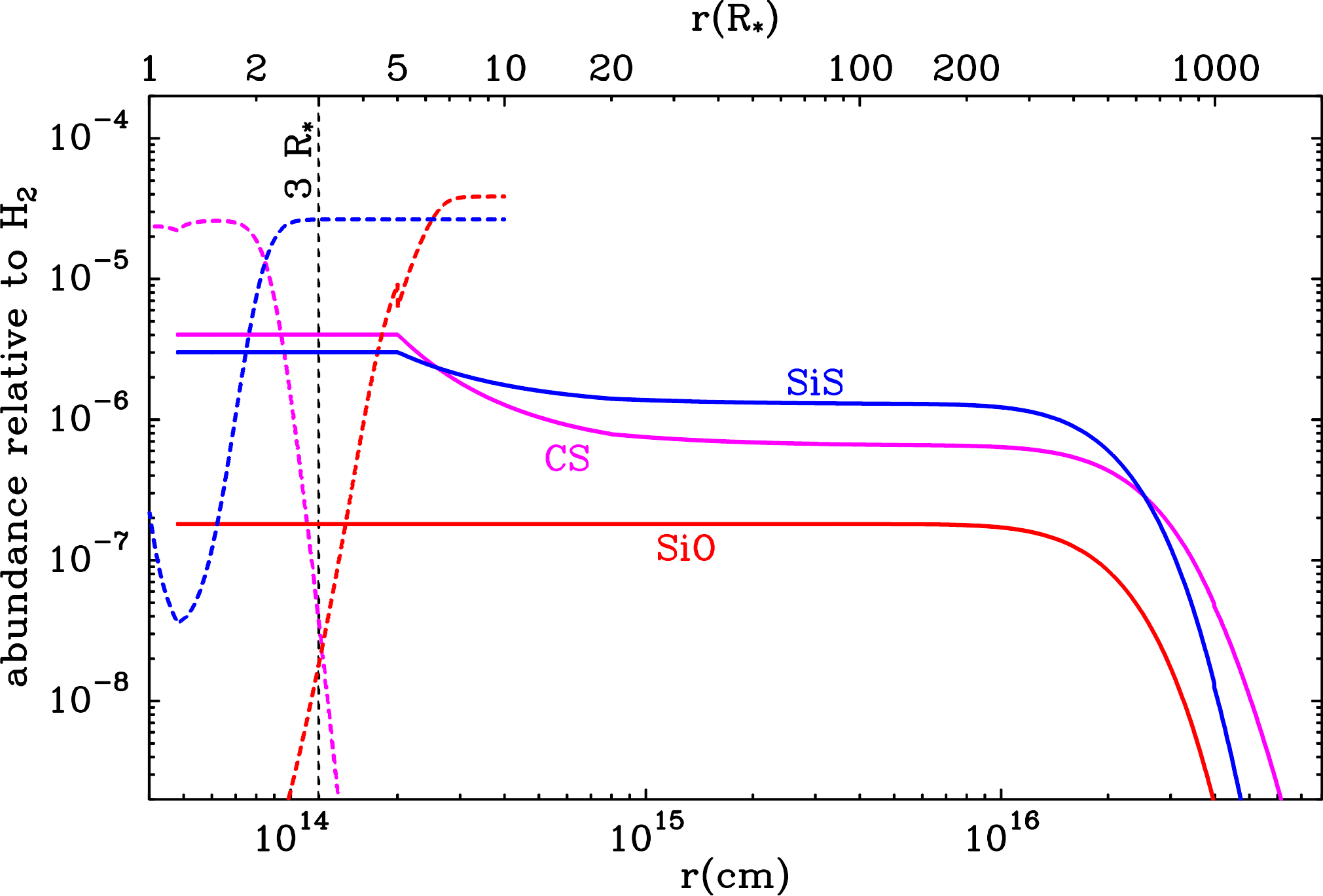}
\caption{Abundances of CS, SiO, and SiS in IRC +10216, as derived
from the radiative transfer model (continuous lines) and as
calculated through thermochemical equilibrium in the innermost
regions of the envelope (dashed lines). A vertical dashed line
indicates the outer boundary where thermochemical equilibrium is
valid ($\sim$3 R$_*$).} \label{fig-abun-cs}
\end{center}
\end{figure}

\begin{figure}
\begin{center}
\includegraphics[angle=0,scale=.43]{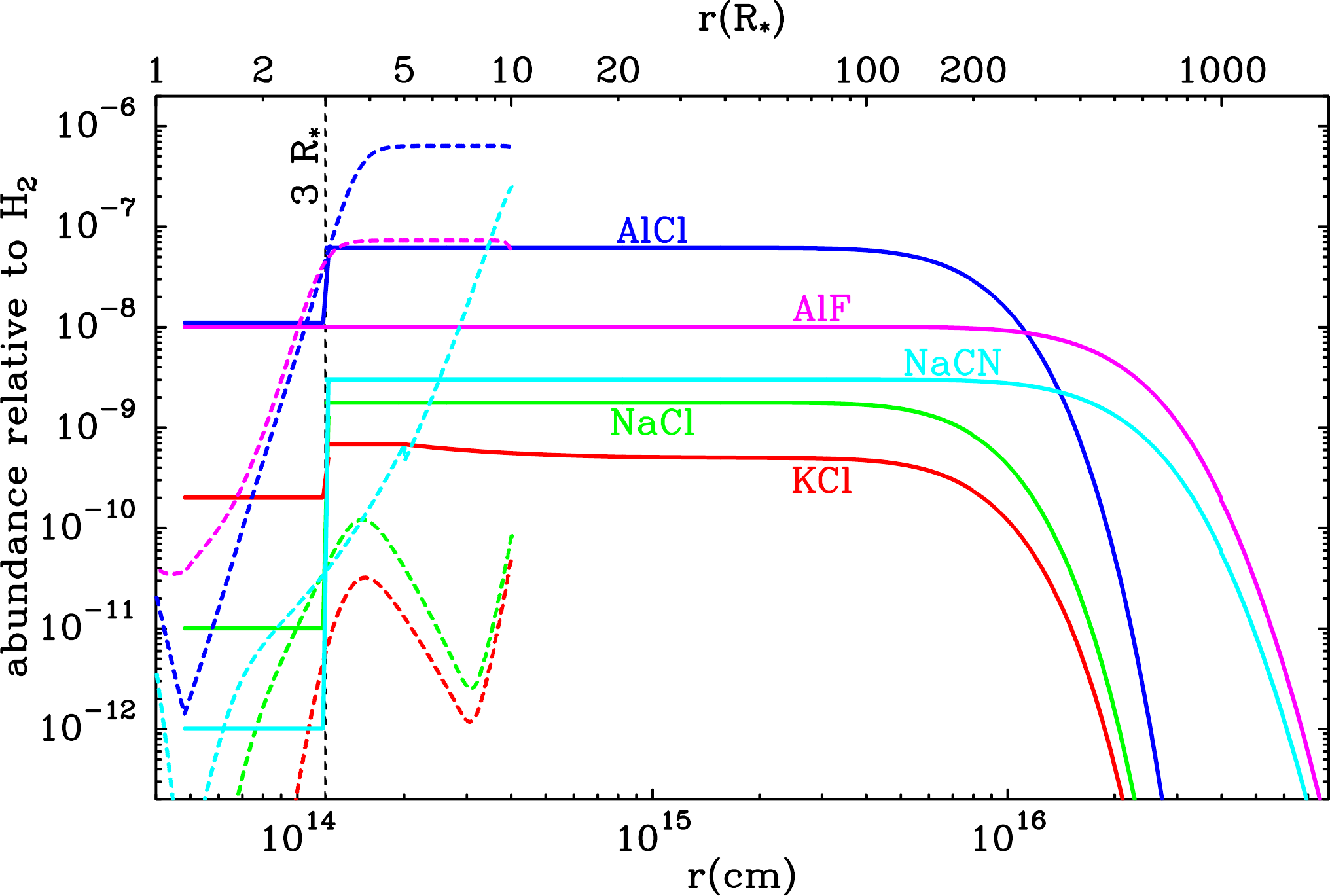}
\caption{Abundances of NaCl, KCl, AlCl (including $^{35}$Cl and
$^{37}$Cl), AlF, and NaCN in IRC +10216, as derived from the
radiative transfer model (continuous lines) and as calculated
through thermochemical equilibrium in the innermost regions of the
envelope (dashed lines). A vertical dashed line indicates the
outer boundary where thermochemical equilibrium is valid ($\sim$3
R$_*$).} \label{fig-abun-nacl}
\end{center}
\end{figure}

A compilation of molecular abundances for the inner layers of IRC
+10216, resulting from this work and from previous studies, is
given in Table~\ref{table-abundances}. The abundance profiles
derived in this study are shown in Fig.~\ref{fig-abun-cs} for CS,
SiO, and SiS, and in Fig.~\ref{fig-abun-nacl} for NaCl, KCl, AlCl,
AlF, and NaCN. For chlorine--containing molecules, the abundances
given in Table~\ref{table-abundances} and shown in
Fig.~\ref{fig-abun-nacl} include both the $^{35}$Cl and $^{37}$Cl
isotopic species, with a $^{35}$Cl/$^{37}$Cl ratio of 2.9, as
derived in Sec.~\ref{subsec-isotopic-ratios}.

In Fig.~\ref{fig-abun-cs} and Fig.~\ref{fig-abun-nacl} we also
show the abundances computed under thermochemical equilibrium,
which is expected to be valid from the stellar photosphere up to a
radius of $\sim$3 R$_*$, from which the decrease in density and
temperature causes an increase in the chemical time scale so that
the abundances experience a quenching effect (see \cite{agu2006}
2006). The thermochemical equilibrium calculations use solar
elemental abundances (\cite{asp2009} 2009), except for carbon
whose abundance is increased over the oxygen value by a factor of
1.5. More details on the thermochemical equilibrium calculations
can be found in \cite{agu2009} (2009). By looking at
Fig.~\ref{fig-abun-cs} and Fig.~\ref{fig-abun-nacl}, we note that
for most of the studied molecules -- a notable exception is CS --
the abundance calculated under thermochemical equilibrium shows a
positive gradient in the 1--3 R$_*$ region, i.e. the abundance
close to the photosphere is much lower than in the surroundings of
the quenching region ($\sim$3 R$_*$). This behavior justifies the
decrease in abundance in the 1--3 R$_*$ region adopted for various
metal--bearing molecules.

For some species (CS, SiS, and KCl) we have found necessary to
consider a negative gradient in the abundance, with a higher value
in the inner layers than in the mid and outer regions, in order to
adequately reproduce the wide range of low-- and high--energy
observed transitions. The scientific rationale for such a gradient
is that the studied molecules contain refractory elements and are
likely to deplete from the gas phase to condense onto dust grains.
Therefore, we consider that the gradient starts at the
condensation radius $R_c$ and continues up to a radius of a few
10$^{15}$ cm. Farther out, at about 10$^{16}$ cm all molecules
start to be photodissociated by the ambient interstellar
ultraviolet field. The observation of a large number of rotational
lines for the different molecules studied, involving levels with a
wide range of energies (including highly excited vibrational
levels) allows us to trace the material, and therefore to
determine the abundance, from the innermost layers out to the
outer envelope.\\

Carbon monosulphide (CS) is found to have an abundance relative to
H$_2$ of 4 $\times$ 10$^{-6}$ in the inner layers, decreasing in
the mid envelope down to a value of 7 $\times$ 10$^{-7}$ at a
radius of 2 $\times$ 10$^{15}$ cm, which serves as input value for
the photochemistry taking place in the outer envelope (see
Fig.~\ref{fig-abun-cs}). Our estimate for the inner layers may be
compared with the results of the thermochemical equilibrium
calculations, which yield a steep abundance gradient around the
quenching region, from 2 $\times$ 10$^{-5}$ at 2 R$_*$ down to 3
$\times$ 10$^{-11}$ at 4 R$_*$. The intersection of our derived
abundance profile with that calculated under thermochemical
equilibrium indicates that the abundance of CS is quenched at a
radius of 2.4 R$_*$. The model built by \cite{wil1998} (1998),
which considers the effect of shocks induced by the stellar
pulsation on the chemistry, predicts abundances somewhat lower in
the inner layers, 2--6 $\times$ 10$^{-7}$ in the 2.7--5 R$_*$
region. Our findings for the inner layers are in perfect agreement
with the previous results of \cite{kea1993} (1993), who derive an
abundance of 4 $\times$ 10$^{-6}$ for the region inner to 17 R$_*$
from observations of ro-vibrational lines in the infrared, but
differ significantly from the estimates made by \cite{hig2000}
(2000), 3--7 $\times$ 10$^{-5}$ at 14 R$_*$, and by \cite{pat2009}
(2009), $>$9.3 $\times$ 10$^{-6}$ within a radius of 7 R$_*$,
which are based on observations of rotational lines in
vibrationally excited states and an analysis involving column
densities and mean excitation temperatures. As concerns the
abundance in the mid and outer envelope, our estimate, 7 $\times$
10$^{-7}$, is somewhat higher than the previous value derived by
\cite{hen1985} (1985), 1.2 $\times$ 10$^{-7}$, based on the
observation of low$-J$ rotational lines within the ground
vibrational state.\\

For silicon monoxide (SiO) we derive an abundance relative to
H$_2$ of 1.8 $\times$ 10$^{-7}$, from the innermost circumstellar
layers up to the photodissociation region. Yet, as in the case of
CS, chemical equilibrium calculations (see Fig.~\ref{fig-abun-cs})
yield a steep gradient near the quenching region, with a value as
low as 3 $\times$ 10$^{-10}$ at 2 R$_*$ increasing up to 10$^{-6}$
at 4 R$_*$. The quenching radius for SiO would be located around
3.6 R$_*$, somewhat farther than that of CS. We find no evidence
of such a low SiO abundance in the innermost layers, perhaps
because only a few vibrationally excited lines tracing this region
enter in our analysis. Calculations based on chemical kinetics
(see model MH in Fig.~3 of \cite{agu2006} 2006) indicate that the
abundance of SiO increases when moving away from the photosphere
due to the reaction between Si and CO, which after a few stellar
radii becomes too slow compared to the dynamical time scale, so
that the abundance of SiO gets quenched. The chemical model with
shocks of \cite{wil1998} (1998) predicts an abundance of 2--7
$\times$ 10$^{-7}$ in the 1.9--5 R$_*$ region, which is in good
agreement with our derived value.

As concerns previous observational studies, one of the most
complete dealing with the abundance of SiO in IRC +10216 is that
made by \cite{sch2006a} (2006a), who used observations of several
rotational lines within the ground vibrational state,
interferometric observations of the $v=0$ $J=5-4$ line, as well as
ro-vibrational lines observed at 8 $\mu$m to derive the abundance
of SiO throughout the envelope. They found an abundance of 3
$\times$ 10$^{-8}$ from the photosphere up to 1.7 $\times$
10$^{14}$ cm, a high abundance component of 1.5 $\times$ 10$^{-6}$
extending up to 4.5 $\times$ 10$^{14}$ cm (which is necessary to
reproduce the ro-vibrational lines observed), and an abundance of
1.7 $\times$ 10$^{-7}$ from this latter radius up to the outer
layers. The main difference with our results is that
\cite{sch2006a} (2006a) find necessary to adopt a high abundance
component in the inner layers. A higher abundance than derived by
us, 8 $\times$ 10$^{-7}$, was also found by \cite{kea1993} (1993)
in their analysis of the observations at 8 $\mu$m. These authors
adopt lower densities than us for the inner layers, they assume
that Eq.~(\ref{eq-density-mass-conservation}) applies down to the
photosphere, and this could explain the higher abundances they
get. On the other hand, the abundance derived by \cite{kea1993}
(1993) for CS is in very good agreement with our value (see
above), in spite of the very different densities adopted for the
inner regions. Given the different nature of the lines used to
trace the abundance in the inner layers, infrared ro-vibrational
lines versus millimeter rotational lines in vibrationally excited
states, it is difficult to explain why the abundances found by
\cite{kea1993} (1993) and by us are different for SiO but not for
CS. A forthcoming study of new high spectral resolution
observations (R$\sim$100,000) of IRC +10216 around 8 $\mu$m
obtained from ground (Fonfr\'ia et al. in preparation) will help
to solve the discrepancy found between our derived SiO abundance
and that found by \cite{kea1993} (1993).

More recently, \cite{dec2010a} (2010a) have observed a large
number of SiO rotational transitions in IRC +10216, some of them
in the $v=1$ state, at low spectral resolution with the SPIRE and
PACS instruments on board \emph{Herschel}. These authors derive an
abundance of 10$^{-7}$ for SiO from the innermost regions up to
the outer layers, in good agreement with our findings.\\

\begin{table}
\caption{Molecular abundances in the inner layers of IRC +10216}
\label{table-abundances} \centering
\begin{tabular}{l@{\hspace{0.8cm}}cccr}
\hline \hline
Molecule & \multicolumn{3}{c}{Abundance relative to H$_2$} & Reference \\
\cline{2-4}
 & 1--5 R$_*$ & $\rightarrow$ & 2 $\times$ 10$^{15}$ cm & \\
\hline
CS         & \multicolumn{1}{c}{4 $\times$ 10$^{-6}$}  & $\rightarrow$     & \multicolumn{1}{r}{7 $\times$ 10$^{-7}$}   & (1) \\
SiO        & \multicolumn{3}{r}{..................................  1.8 $\times$ 10$^{-7}$} & (1) \\
SiS        & \multicolumn{1}{c}{3 $\times$ 10$^{-6}$}  & $\rightarrow$     & \multicolumn{1}{r}{1.3 $\times$ 10$^{-6}$} & (1) \\
NaCl$^a$   & \multicolumn{3}{r}{.................................. 1.8 $\times$ 10$^{-9}$}  & (1) \\
KCl$^a$    & \multicolumn{1}{c}{7 $\times$ 10$^{-10}$} & $\rightarrow$     & \multicolumn{1}{r}{5 $\times$ 10$^{-10}$}  & (1) \\
AlCl$^a$   & \multicolumn{3}{r}{..................................... 7 $\times$ 10$^{-8}$} & (1) \\
AlF        & \multicolumn{3}{r}{..................................... 1 $\times$ 10$^{-8}$} & (1) \\
NaCN$^a$   & \multicolumn{3}{r}{..................................... 3 $\times$ 10$^{-9}$} & (1) \\
\hline
CO         & \multicolumn{3}{r}{..................................... 6 $\times$ 10$^{-4}$} & (1) \\
C$_2$H$_2$ & \multicolumn{3}{r}{..................................... 8 $\times$ 10$^{-5}$} & (2) \\
HCN        & \multicolumn{3}{r}{..................................... 2 $\times$ 10$^{-5}$} & (2) \\
CH$_4$     & \multicolumn{3}{r}{..................................  3.5 $\times$ 10$^{-6}$} & (3) \\
NH$_3$     & \multicolumn{3}{r}{..................................... 2 $\times$ 10$^{-6}$} & (4) \\
SiH$_4$    & \multicolumn{3}{r}{..................................  2.2 $\times$ 10$^{-7}$} & (3) \\
SiC$_2$    & \multicolumn{3}{r}{..................................... 2 $\times$ 10$^{-7}$} & (5) \\
H$_2$O     & \multicolumn{3}{r}{..................................... 1 $\times$ 10$^{-7}$} & (6) \\
HCl        & \multicolumn{3}{r}{..................................... 1 $\times$ 10$^{-7}$} & (7) \\
HCP        & \multicolumn{3}{r}{..................................  2.5 $\times$ 10$^{-8}$} & (8) \\
C$_2$H$_4$ & \multicolumn{3}{r}{..................................... 2 $\times$ 10$^{-8}$} & (9) \\
HF         & \multicolumn{3}{r}{..................................... 8 $\times$ 10$^{-9}$} & (7) \\
PH$_3$     & \multicolumn{3}{r}{..................................... 8 $\times$ 10$^{-9}$} & (10) \\
H$_2$S     & \multicolumn{3}{r}{..................................... 4 $\times$ 10$^{-9}$} & (11) \\
\hline
\end{tabular}
\tablenotec{\\
\textsc{Notes:} $^a$~Abundance in the 1--3 R$_*$ region is reduced
to 1 $\times$ 10$^{-11}$ (NaCl), 2 $\times$ 10$^{-10}$ (KCl), 1
$\times$ 10$^{-8}$ (AlCl), and 1 $\times$ 10$^{-12}$ (NaCN).\\
\textsc{References:} (1) this study (NaCl, KCl, and AlCl
abundances include both $^{35}$Cl and $^{37}$Cl isotopomers); (2)
\cite{fon2008} (2008); (3) \cite{kea1993} (1993); (4)
\cite{has2006} (2006); (5) \cite{cer2010} (2010); (6)
\cite{dec2010b} (2010b); (7) \cite{agu2011} (2011); (8) based on
\cite{agu2007} (2007); (9) \cite{gol1987} (1987); (10) based on
\cite{agu2008} (2008); (11) derived from the line $1_{\rm
1,0}-1_{\rm 0,1}$ observed by \cite{cer2000} (2000).}
\end{table}

According to our best model, silicon monosulphide (SiS) is present
in the inner layers with an abundance relative to H$_2$ of 3
$\times$ 10$^{-6}$, decreasing in the mid envelope down to a value
of 1.3 $\times$ 10$^{-6}$ at a radius of 2 $\times$ 10$^{15}$ cm.
As shown in Fig.~\ref{fig-abun-cs}, the thermochemical equilibrium
abundance of SiS shows a steep gradient, rising from 10$^{-7}$ at
the photosphere up to a maximum value of 2.6 $\times$ 10$^{-5}$
beyond 2.5 R$_*$. Our derived abundance is substantially lower
than this latter value, which points towards the SiS abundance
being quenched in a region inner to 2.5 R$_*$, concretely at 1.9
R$_*$, as deduced from the intersection of the continuous and
dashed lines in Fig.~\ref{fig-abun-cs}. Calculations based on
chemical kinetics (\cite{wil1998} 1998; \cite{agu2006} 2006)
result in SiS abundances of 2--3 $\times$ 10$^{-5}$ in the regions
inner to 5 R$_*$, about one order of magnitude larger than the
value we find. These models are, however, affected by
uncertainties due to the poor knowledge of rate constants for
reactions involving Si-- and S--bearing species.

As concerns previous observational studies, there is a general
consensus on the presence of a gradient in the abundance of SiS,
with reported values in the range 10$^{-6}$--10$^{-5}$ in the
inner layers ($>$6.5 $\times$ 10$^{-7}$ by \cite{tur1987} 1987;
7.5 $\times$ 10$^{-6}$ by \cite{bie1989} 1989; 4.3 $\times$
10$^{-5}$ close to the stellar surface decreasing down to 4.3
$\times$ 10$^{-6}$ at 12 R$_*$ by \cite{boy1994} 1994) and
reported values in the range 10$^{-7}$--10$^{-6}$ in the outer
envelope (2.4 $\times$ 10$^{-7}$ by \cite{sah1984} 1984; 5
$\times$ 10$^{-7}$ by \cite{ngu1984} 1984; 5 $\times$ 10$^{-7}$ by
\cite{hen1983} 1983; 1.5--4 $\times$ 10$^{-7}$ by \cite{hen1985}
1985; 6.5 $\times$ 10$^{-7}$ by \cite{bie1989} 1989; 1.4 $\times$
10$^{-6}$ by \cite{sch2007} 2007). The more recent study of
\cite{dec2010a} (2010a), based on observations with SPIRE and PACS
that are particularly sensitive to the warm inner envelope,
reports a SiS abundance of 4 $\times$ 10$^{-6}$, in good agreement
with our finding for the inner layers.\\

The most abundant sulfur--bearing molecules in IRC +10216 are
therefore CS and SiS, H$_2$S being a minor species (see
Table~\ref{table-abundances}). Assuming a solar elemental
abundance, we conclude that 27 \% of the total available sulfur is
locked in gas phase CS and SiS molecules in the inner layers. The
fraction decreases down to 7.6 \% in the outer envelope, an
evidence of sulfur depletion onto dust grains in the dust
formation region. A fraction of the remaining sulfur could be in
atomic form, but it is likely that the bulk of sulfur is in the
form of solid condensates such as MgS (\cite{goe1985} 1985).
Several silicon--bearing molecules such as SiS, SiO, SiC$_2$, and
SiH$_4$ are present in the inner envelope, all them locking up to
5.6 \% of the total available silicon. This small fraction,
together with the fact that SiS (the most abundant Si--containing
molecule) gets depleted by a factor of 2.3 after the dust
condensation region, indicates that silicon massively condenses
into dust grains, mainly forming SiC (\cite{tre1974} 1974). We see
no evidence of SiO depletion at the dust formation region in our
SiO abundance profile. One would expect a depletion similar to
that of SiS, silicon being a refractory element that is thought to
easily condense onto dust grains in the envelopes of AGB stars
(\cite{sch2006b} 2006b). This may result from our small sample of
vibrationally excited SiO lines, which limits our sensitivity to
the innermost layers.\\

The metal halides NaCl, KCl, AlCl, and AlF, and the cyanide NaCN
are among the first metal--containing molecules detected in space
(\cite{cer1987} 1987; \cite{tur1994} 1994). Metal--bearing
molecules are rarely observed in the gas phase of the interstellar
medium due to their large refractory character that make them to
easily form solid condensates. Their presence in IRC +10216 is
largely caused by the high temperatures and densities prevailing
close to the star, where, under thermochemical equilibrium, a good
fraction of metals are in the gas phase forming this kind of
molecules.

The derived abundances relative to H$_2$ in the inner layers are
1.8 $\times$ 10$^{-9}$ for NaCl, 7 $\times$ 10$^{-10}$ for KCl, 7
$\times$ 10$^{-8}$ for AlCl, 10$^{-8}$ for AlF, and 3 $\times$
10$^{-9}$ for NaCN. For some of these molecules the abundance has
been decreased in the 1--3 R$_*$ region, in order to better
account for the observed line profiles. The rationale for this
abundance modification is related to the thermochemical
equilibrium calculations, which for all these species show a
strong abundance gradient in this region, with low values close to
the stellar surface and higher values around the quenching region
(see Fig.~\ref{fig-abun-nacl}). For NaCl and KCl, the abundance
decrease in this 1--3 R$_*$ region improves significantly the
agreement between calculated and observed line profiles,
indicating that these species are likely depleted close to the
photosphere by at least a factor of $\sim$100 for NaCl and
$\sim$2--3 for KCl. For AlCl and NaCN, the abundance decrease
adopted in the 1--3 R$_*$ region affects only slightly the
profiles of the observed lines, while for AlF no abundance
modification has been adopted since line profiles are almost not
affected. The observed lines of AlCl, AlF, and NaCN are therefore
not sensitive enough to the abundance in this innermost region,
and it cannot be excluded that these molecules have an abundance
as low as predicted by TE close to the star. Vibrationally excited
lines should allow to better probe this region.

The abundances calculated under thermochemical equilibrium are in
reasonable agreement with our derived values, although some
remarks should be made. For the metal chlorides NaCl and KCl, the
derived abundances are higher by about one order of magnitude than
the maximum values given by thermochemical equilibrium in the
surroundings of the quenching region (see
Fig.~\ref{fig-abun-nacl}), something that could be indicative of
non-equilibrium processes being at work to enhance the abundances
of NaCl and KCl. In the case of AlCl and AlF, the derived
abundances are in good agreement with the thermochemical
equilibrium calculations, pointing towards the abundance of these
two molecules being quenched at a radius of 2.6 R$_*$. For NaCN,
the analysis is not so obvious as the abundance calculated under
thermochemical equilibrium shows a steep positive gradient that
reaches the derived abundance of 3 $\times$ 10$^{-9}$ at a radius
of $\sim$6 R$_*$, somewhat farther than the typical quenching
region located around 3 R$_*$. The lack of chemical kinetics data
on reactions involving NaCN makes it difficult to evaluate whether
such a large quenching radius is still possible or not.

Metals are refractory elements, so that we expect
metal--containing molecules to deplete from the gas phase to
incorporate into dust grains in the condensation region.
Surprisingly, we find no evidence for such depletion except for
KCl, whose abundance is found to slightly decrease from its value
of 5 $\times$ 10$^{-10}$ in the inner layers down to 3.7 $\times$
10$^{-10}$ at a radius of 2 $\times$ 10$^{15}$ cm. It therefore
seems that the presence of metals in the gas phase is not just
restricted to the warm inner circumstellar regions, but also to
the cool outer regions where a non negligible fraction survives in
the gas phase. Further evidence to this respect comes from the
observation of other metal--bearing molecules in the outer layers
of IRC +10216, such as MgNC (\cite{gue1993} 1993), MgCN
(\cite{ziu1995} 1995), AlNC (\cite{ziu2002} 2002) KCN
(\cite{pul2010} 2010), and FeCN (\cite{zac2011} 2011), and from
the remarkable detection of gas phase atomic metals in the outer
envelope of IRC +10216 by \cite{mau2010} (2010). As already
pointed out by these latter authors, metals must be mostly in
solid state in the outer layers of IRC +10216 since gas phase
species do only account for a relatively small fraction of the
total available elemental abundance, 24 \% as Na{\footnotesize I}
and Na{\footnotesize II} and 0.14 \% as NaCl and NaCN in the case
of sodium, 5 \% as K{\footnotesize I} and K{\footnotesize II} and
0.5 \% as KCl and KCN as concerns potassium, and 1.4 \% as AlCl,
AlF, and AlNC in the case of aluminium. These small percentages
are however remarkable, given the low temperatures prevailing in
the outer layers, and are high enough to allow for a metal--based
gas phase chemistry to take place. In the inner layers the
situation should not be very different, as inferred from the
derived molecular abundances, with metals being mostly in the form
of solid condensates (only beyond the condensation radius) and gas
phase atoms, and a small but sizable fraction being locked in gas
phase molecules.

\subsection{Isotopic ratios} \label{subsec-isotopic-ratios}

For some of the studied molecules, a good number of transitions of
rare isotopomers have been observed (see
Tables~\ref{table-cs-line-param}--\ref{table-alcl-line-param}),
allowing us to determine the corresponding isotopic abundance
ratios. For each of these rare isotopomers we have adopted the
abundance profile derived for the main isotopomer, decreased by
the corresponding isotopic ratio, which is varied until the line
profiles calculated with the radiative transfer model reproduce
the observed ones. As compared with the previous method used to
derive isotopic ratios from line intensity ratios (e.g.
\cite{cer2000} 2000; \cite{kah2000} 2000), the method used here,
based on LVG radiative transfer calculations, can better, although
not completely, deal with optical depth effects present in the
lines of some main isotopomers. Such effects can be completely
avoided by using optically thin lines of doubly substituted
isotopic species, but such lines are weak and are observed with
low signal-to-noise ratios, which may introduce large
uncertainties in the derived isotopic ratios. Here, we have
therefore focused on single substituted isotopomers.

\begin{table}
\caption{Isotopic ratios in IRC +10216}
\label{table-isotopic-ratios} \centering
\begin{tabular}{l@{\hspace{1.3cm}}r@{\hspace{0.8cm}}r@{\hspace{0.8cm}}r}
\hline
Isotopic ratio      & Value             & Literature            & Solar$^g$ \\
\hline
$^{12}$C/$^{13}$C   & 35 $\pm$ 3.5$^a$  & 45 $\pm$ 3$^e$        & 89.4 \\
$^{32}$S/$^{34}$S   & 22 $\pm$ 2.5$^b$  & 21.8 $\pm$ 2.6$^e$    & 22.1 \\
$^{32}$S/$^{33}$S   & 112 $\pm$ 12$^b$  & 121 $\pm$ 21$^e$      & 124.9 \\
$^{28}$Si/$^{29}$Si & 18 $\pm$ 2$^c$    & $>$15.4$^e$           & 19.7 \\
$^{28}$Si/$^{30}$Si & 27 $\pm$ 3.0$^c$  & $>$22.3$^e$           & 29.9 \\
$^{35}$Cl/$^{37}$Cl & 2.9 $\pm$ 0.3$^d$ & 2.3 $\pm$ 0.24$^e$    & 3.13 \\
                    &                   & 3.3 $\pm$ 0.3$^f$     & \\
\hline
\end{tabular}
\tablenoted{$^a$~From CS; $^b$~from SiS; $^c$~from SiO and SiS;
$^d$~from NaCl, KCl, and AlCl; $^e$~\cite{kah2000} (2000) and
\cite{cer2000} (2000); $^f$~\cite{agu2011} (2011);
$^g$~\cite{asp2009} (2009).}
\end{table}

The derived isotopic ratios are given in
Table~\ref{table-isotopic-ratios}. Under the reasonable assumption
of a similar excitation and spatial extent for two isotopomers,
most of the errors that usually affect the determination of an
abundance, cancel when deriving an isotopic ratio (see
\cite{kah2000} 2000). The estimated uncertainties given in
Table~\ref{table-isotopic-ratios}, therefore, neglect any error
coming from the model and are simply based on the signal-to-noise
ratio of the observed lines (of main and rare isotopomers), and on
the sensitivity of the calculated line profiles to the adopted
isotopic ratio.

We find an elemental $^{12}$C/$^{13}$C ratio of 35 $\pm$ 3.5 from
the observed lines of $^{12}$CS and $^{13}$CS (see
Table~\ref{table-cs-line-param}), somewhat lower than the value of
45 $\pm$ 3 derived from intensity ratios of optically thin lines
of SiC$_2$ and CS (\cite{cer2000} 2000). In the case of the sulfur
isotope $^{32}$S, we derive values of C$^{32}$S/C$^{34}$S = 16
$\pm$ 2.5 and Si$^{32}$S/Si$^{34}$S = 22 $\pm$ 2.5. The different
values should result, rather than from $^{32}$S isotopic
fractionation which is unlikely in warm regions, from optical
depth effects that are more important for CS than for SiS. We thus
adopt the Si$^{32}$S/Si$^{34}$S isotopic ratio as a better proxy
of the elemental $^{32}$S/$^{34}$S abundance ratio. A similar
situation is found for the sulfur isotope $^{33}$S, for which we
derive C$^{32}$S/C$^{33}$S = 83 $\pm$ 11 and Si$^{32}$S/Si$^{33}$S
= 112 $\pm$ 12, the latter value being a better estimation of the
elemental $^{32}$S/$^{33}$S. The isotopic ratios found for carbon
and sulfur are similar or slightly smaller than those previously
derived by \cite{cer2000} (2000) from intensity ratios of
optically thin lines. The fact that the $^{12}$C/$^{13}$C and
$^{32}$S/$^{33}$S ratios derived here are somewhat smaller than
those found by \cite{cer2000} (2000) probably indicates that our
results are affected to some extent by optical depth effects. The
usefulness of the method utilized here is clear in cases where
optically thin lines (e.g. of doubly substituted isotopomers) are
close or below the detection limit of the observations. This
occurs in the cases of the $^{28}$Si/$^{29}$Si and
$^{28}$Si/$^{30}$Si ratios, for which we find values (see
Table~\ref{table-isotopic-ratios}) consistent with the lower
limits given by \cite{cer2000} (2000).

The $^{35}$Cl/$^{37}$Cl isotopic ratio is found to be 2.9 $\pm$
0.3 from the lines of NaCl, KCl, and AlCl, which is in between the
value of 2.3 $\pm$ 0.24 found by \cite{kah2000} (2000) from
intensity ratios of some lines of NaCl and AlCl, and the value of
3.3 $\pm$ 0.3 derived by \cite{agu2011} (2011) from the modeling
of the $J=1-0$ to $J=3-2$ lines of HCl. The low
$^{35}$Cl/$^{37}$Cl ratio found by \cite{kah2000} (2000) indicates
that lines of NaCl and AlCl could be affected by optical depth
effects, so that their calculated value based on line intensity
ratios would be a lower limit to the true isotopic ratio. Finally,
as concerns the $^{16}$O/$^{17}$O and $^{16}$O/$^{18}$O ratios, we
have not attempted to determine them due to the low
signal-to-noise ratios of the observed Si$^{17}$O and Si$^{18}$O
lines (see Table~\ref{table-sio-line-param}). We nevertheless find
that the observed lines of these two rare isotopomers are
consistent with the oxygen isotopic ratios $^{16}$O/$^{17}$O = 840
and $^{16}$O/$^{18}$O = 1260 given by \cite{kah2000} (2000).

The isotopic ratios derived here, although systematically smaller,
are very close to the solar values, to the exception of
$^{12}$C/$^{13}$C (see Table~\ref{table-isotopic-ratios}). since a
bias towards smaller ratios is expected, due to the residual
optical depth effects, we conclude that the elemental isotopic
ratios $^{32}$S/$^{34}$S, $^{32}$S/$^{33}$S, $^{28}$Si/$^{29}$Si,
$^{28}$Si/$^{30}$Si, and $^{35}$Cl/$^{37}$Cl in IRC +10216 are
compatible with the solar values. This result is not unexpected
since the central AGB star is not massive enough to form much S,
Si, or Cl. The lower $^{12}$C/$^{13}$C ratio found in IRC +10216,
as compared with than in the Sun, is known to result from CNO
cycle nuclear processing in the interior of the star
(\cite{kah2000} 2000).

\section{Summary}

Astronomical observations of the carbon star envelope IRC +10216
extending over the full frequency coverage of the IRAM 30-m
telescope (80--357.5 GHz) have been analyzed to obtain accurate
abundances for several molecules formed in the inner layers: CS,
SiO, and SiS, the metal halides NaCl, KCl, AlCl, and AlF, and the
metal cyanide NaCN. The observation of a large number of
rotational transitions covering a wide range of energy levels,
including highly excited vibrational states, allow us to derive
abundances from the innermost layers out to the outer regions
where molecules are photodissociated by interstellar ultraviolet
photons. For some molecules, noticeably CS and SiS, we find that
abundances are depleted from the inner to the outer layers,
implying that they contribute to the formation of dust. The amount
of sulfur and silicon locked up in gas phase molecules in the
inner layers, 27 \% and 5.6 \% respectively, indicates that a
major fraction of these elements is in the form of solid
condensates, most likely as MgS and SiC, in IRC +10216's envelope.
Metal--bearing molecules lock a relatively small fraction (a few
percent) of metals. Nevertheless, we find that NaCl, KCl, AlCl,
AlF, and NaCN, which all contain refractory elements, seem to
maintain a surprisingly high abundance in the cold outer layers,
where they can participate, together with metallic atoms, in a
rich gas phase chemistry. The lines of rare isotopomers of CS,
SiO, SiS, and some metal compounds have been analyzed, and we
derive abundance ratios for rare isotopes of carbon, sulfur,
silicon, and chlorine. The values obtained are in good agreement
with previously reported values. They are consistent with the
solar values, except for the $^{12}$C/$^{13}$C ratio, which is
more than twice smaller in IRC +10216 than in the Sun, in
agreement with previous findings.

\begin{acknowledgements}

We acknowledge the referee, H. Olofsson, for a constructive
report. M.A., J.C., and F.D. thank the Spanish MICINN for funding
support through the grant AYA2009-07304 and the ASTROMOL
Consolider project CSD2009-00038. During this study M.A. has been
supported by Spanish MEC grant AP2003-4619, Spanish MICINN
projects PIE 200750I024 and PIE 200750I028, \emph{Marie Curie
Intra-European Individual Fellowship} 235753, and ERC project
209622 E$_3$ARTHs. J.P.F. has been supported by the UNAM through a
postdoctoral fellowship.

\end{acknowledgements}

\end{document}